\documentclass[acmtog,nonacm,balance=True]{acmart}
\acmSubmissionID{495}
\renewcommand\footnotetextcopyrightpermission[1]{}
\renewcommand\footnotetextauthorsaddresses[1]{}

\usepackage{booktabs} % For formal tables

\usepackage{bm}
\usepackage{makecell}
\usepackage{xspace}
\usepackage{enumitem}

\usepackage{multirow} 
\AtEndPreamble{
    \usepackage[capitalize]{cleveref}
    \crefname{section}{Sec.}{Secs.}
    \Crefname{section}{Section}{Sections}
    \Crefname{table}{Table}{Tables}
    \crefname{table}{Tab.}{Tabs.}
}

\def\eg{e.g.,~}

\def\etc{etc.~}

\usepackage[ruled,noend]{algorithm2e} % For algorithms

\SetKwComment{Comment}{\color{green!50!black}// }{}

\newcommand{\var}{\texttt}
\newcommand{\FuncCall}[2]{\texttt{\bfseries #1(#2)}}
\SetKwProg{Function}{function}{}{}

\newcommand{\methodA}{Far-field}

\begin{document}
% Title portion
\title{8DNA: 8D Neural Asset Light Transport by Distribution Learning}

\author{Liwen Wu}
\orcid{0009-0007-2773-2032}
\affiliation{
    \institution{University of California San Diego}
    \country{USA}
}
\email{liw026@ucsd.edu}

\author{Haolin Lu}
\orcid{0009-0008-2595-2493}
\affiliation{
    \institution{University of California San Diego}
    \country{USA}
}
\email{hal128@ucsd.edu}
 
\author{Bing Xu}
\orcid{0009-0005-7359-8570}
\affiliation{
    \institution{University of California San Diego}
    \country{USA}
}
\affiliation{
    \institution{NVIDIA}
    \country{USA}
}
\email{b4xu@ucsd.edu}

\author{Miloš Hašan}
\orcid{0000-0003-3808-6092}
\affiliation{
    \institution{NVIDIA}
    \country{USA}
}
\email{milos.hasan@gmail.com}

\author{Ravi Ramamoorthi}
\orcid{0000-0003-3993-5789}
\affiliation{
    \institution{University of California San Diego}
    \country{USA}
}
\email{ravir@cs.ucsd.edu}

\begin{abstract}
High-fidelity 3D assets exhibit intriguing global illumination effects like subsurface scattering, glossy interreflections, and fine-scale fiber scatterings, which often involve long scattering paths that are expensive to simulate.
We introduce 8D neural assets (8DNA) to pre-bake these light transport effects 
into neural representations. 
Unlike prior methods that assume far-field lighting
and precompute light transport into 6D functions,
8DNA learns the full 8D light transport, enabling accurate rendering under near-field illumination.
Our training leverages a distribution-learning formulation that learns light transport from forward path-traced samples, which produces less optimization variance with lower training budget than the prior regression-based approaches.
Experiments show our 8DNA rendering closely matches path-traced results under various scene configurations,
yet it achieves improved variance reduction and fast inference speeds on challenging assets.
\end{abstract}

\begin{teaserfigure}
    \centering
    \includegraphics[width=0.99\linewidth]{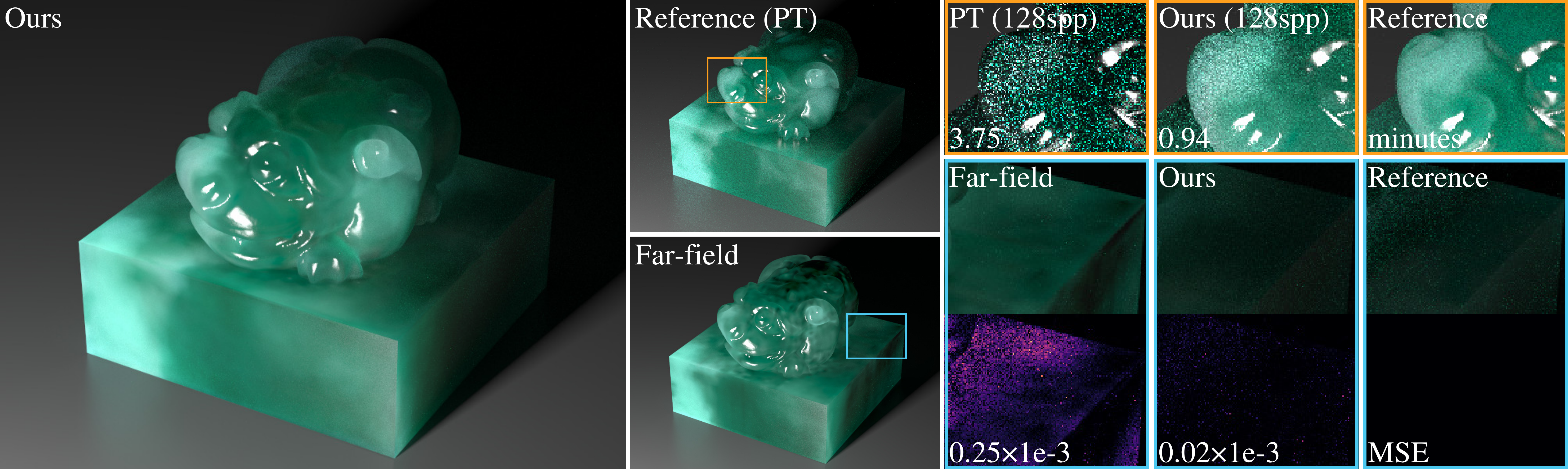}
    \caption{\textbf{Our method vs. the baselines.}
Our method pre-bakes neural assets with complex light transport---including any types of scatterings from volumes to surfaces---that can be imported between renderers for physically based rendering.
This yields much lower rendering variance (insets on top) and faster inference speed (numbers in minutes) than simulating the light transport online with standard path tracing (PT).
In contrast to previous regression-based pre-baking (far-field) that is limited to a far-field approximation of the light transport,
our representation learns the distribution of full 8D light transport from path-traced samples, correctly reproducing the occlusion effects on the back of the seal (insets at bottom).
%Standard path tracing (PT) can be inefficient at rendering complex assets like
%the jade volume enclosed by a dielectric boundary.
%Not only is the variance reduction challenging (first inset on top) but also the long scattering paths are expensive to simulate (numbers in minutes).
%Our method pre-bakes the full internal light transport of the seal---including any types of scatterings from volumes to surfaces---yielding much lower variance and faster inference speed (second inset on top).
%In contrast to previous regression-based pre-baking (far-field) that is limited to a far-field approximation of the light transport,
%our representation learns the distribution of full 8D light transport from path-traced samples, correctly reproducing the occlusion effects on the back of the seal (insets at bottom).
    }
    \label{fig:1-teaser}
\end{teaserfigure}
\maketitle

\newcommand{\xxo}{\mathbf{x}_o}
\newcommand{\xxi}{\mathbf{x}_i}
\newcommand{\omegao}{\bm{\omega}_o}
\newcommand{\omegai}{\bm{\omega}_i}

\newcommand{\bo}{\bm{\omega}}
\newcommand{\bx}{\mathbf{x}}

\newcommand{\capl}{\mathbf{L}}
\newcommand{\capf}{\mathbf{F}}
\newcommand{\fnear}{\mathbf{F}'}

\newcommand{\surface}{\mathcal{M}}
\newcommand{\sphere}{\mathcal{S}}

\section{Introduction}
\label{sec:introduction}
Real-world 3D objects often exhibit interesting and complex light transport effects,
such as multiple scattering within media enclosed by refractive boundaries, glossy interreflections, fine-scale fiber scattering in hair, fur or fabrics, and more. 
These effects are challenging to render in digital 3D assets: they often involve long light paths that are non-trivial to trace from either the camera or the light source, making them expensive to simulate with standard Monte Carlo methods. Furthermore, the appearance models responsible for the above effects can be non-trivial to consistently implement and precisely match between different rendering systems.

Recent work~\cite{mullia2024rna,tg2024neupress} has explored a promising alternative: precompute the light transport within such 3D assets into a neural representation, which can be rendered more efficiently, and transferred between renderers more easily. These methods propose to bake all light paths within a 3D asset into a relightable neural network, which can be evaluated for any point on the asset and any view and light direction.
However, such parameterizations are 6-dimensional: they fundamentally assume distant illumination. While they can still be approximately used under near-field lighting, we show that this can lead to inaccuracy.

The true light transport operator describing light paths through the 3D asset is an 8D function $\capf(\xxo,\omegao,\xxi,\omegai)$: it depends on both outgoing (camera) direction $\omegao$ and position $\xxo$, as well as their incoming (light) counterparts $\omegai$ and $\xxi$. This 8D function is not only challenging to compress, but its queries are non-trivial to estimate with low variance, especially in the most interesting cases: appearances that feature highly specular surface or fiber interactions, and/or forward-scattering phase functions.
%Therefore, it is difficult to train the full 8D transport with a simple regression loss used for the 6D functions~\cite{mullia2024rna,tg2024neupress}.
Therefore, unlike the 6D functions in ~\citet{mullia2024rna} and \citet{tg2024neupress}, it is difficult to train the full 8D transport with a simple regression loss.

%In this paper, we propose to recover the full 8D light transport from forward path samples through the asset in a distribution-learning framework.
In this paper, we propose to recover the full 8D light transport in a distribution-learning framework.
From the forward path samples through the asset,
we estimate the global scattering distribution using a normalizing flow and predict the survival probability (albedo term), together giving a natural decomposition of the global light transport distribution $\capf$.

%We show that, 
In contrast to the challenging direct evaluations of $\capf$,
sampling $\capf$ is as easy as standard path tracing through the asset without next-event estimation. 
%it is comparatively easy to sample forward light paths following the distribution of  $\capf$. 
%A generative neural network such as normalizing flow is used to estimate the global scattering distribution from these path samples.
%A neural network based on normalizing flows or flow matching, similar to architectures used in recent generative models, estimates the global scattering distribution from these path samples.
%The sampling probability distribution induced by $\capf$ is not sufficient to recover $\capf$ itself; however, we show that it is possible to learn a survival probability (albedo term) that scales the probability density values to the corresponding evaluations of $\capf$ itself.
%The efficiency of this sampling approach 
Such efficiency
allows our model to train faster than regression-based baselines while better capturing 8D light transport including accurate near-field illumination.
As a further benefit, the learned representation can be evaluated \emph{and} importance-sampled, with the sampling distribution by construction precisely proportional to the transport.
Similar arguments are made in concurrent work on precomputing multiple scattering from microgeometry~\cite{li2025puresample};
however, this work does not target 3D assets and still assumes 6D far-field light transport.

Figure~\ref{fig:1-teaser} demonstrates that our 8D neural asset (8DNA) successfully reconstructs the global illumination of the jade seal lit by a nearby area emitter: a challenging configuration for the baselines.
Our representation is particularly effective for objects with long scattering paths,
such as translucent assets with complex interiors,
achieving noticeable variance reduction compared to standard path tracing (\cref{sec:experiments}).
In summary, our contributions are:
\begin{itemize}
    %\item We introduce 8DNA, an
    \item An efficient and accurate 8D neural asset representation of arbitrary 3D assets with complex internal light transport.
    %\item We precompute the 8DNA representation using
    \item A forward sampling training framework of our representation,  whose implementation complexity and training cost are as low as standard forward path tracing without next-event estimation.
    %\item We show that
    \item Results showing 8DNA assets rendered within a final scene at an accuracy closely matching ground truth, without tracing long light paths or needing to implement the original light scattering models in the deployment renderer.
\end{itemize}

\section{Preliminaries and Related Work}
\label{sec:background}

\paragraph{8D near-field light transport.} 
Recall our goal of precomputing global light transport within a 3D asset consisting of surfaces, volumes, fibers, etc.
We assume the asset is non-emissive and bounded by a 2D surface $\surface$ (support for volumes without boundary is discussed in \cref{subsec:4-ablation}).
For such assets,
light scattering can be encapsulated by a light transport function
%(weighted by clamped cosine)
$\capf(\xxo,\omegao,\xxi,\omegai)$ that maps incident radiance $\capl_i$ on any boundary point to outgoing radiance $\capl_o$ at another boundary point;
integrations are taken over each variable's domain of definition ($\xxi\in\surface$, $\omegai\in$ sphere) unless specified:
\begin{equation}
%\begin{split}
    \capl_o(\xxo,\omegao; \capl_i) = 
    %\int_{\sphere} \int_{\surface}
    \iint
    \capf(\xxo,\omegao,\xxi,\omegai)\capl_i(\xxi,\omegai)\mathrm{d}\xxi\mathrm{d}\omegai.
    %&= \capl_{pt}(\xxo,\omegao,\capl_i)
%\end{split}
\label{eq:2-glte}
\end{equation}%
%Note that $\capl_i$ denotes incident radiance from the rest of the scene (excluding the internal multiple-scattering within the asset),
%We use $\capl_o(\xxo,\omegao; \capl_i)$ to denote the dependence of the outgoing on the incoming radiance,
Note that $\capl_i$ excludes the internal multiple-scattering within the asset,
and we will use the fact that $\capl_o$ is computable by forward path tracing given any $\capl_i$.
%Note that a clamped cosine term at $\xxi$ is included in $\capf$.
Our objective is to learn the light transport $\capf$ for a given asset, in the form of a neural representation, to enable efficient re-rendering under arbitrary illumination $\capl_i$.
%We will use the fact that $\capl_o$ is computable by forward path tracing given any $\capl_i$, by learning the sampling distribution of such a path tracing process.

%is defined by the rendering equation~\cite{kajiya1986rendering,kajiya1984ray}:
% \input{equations/2-vlte}%
% \lw{Eq. 1 can be removed now as long as we keep a notation to the path-tracing realization of Eq. 2.}
% $\capl_o$ is iteratively path-traced within the asset to account for internal multiple-scattering,

% Note this is different from their usual conversions that $\capl_i$ equals some $\capl_o$ elsewhere in the opposite direction.

%\lw{I assume we will avoid calling our model a BSSDF, I will update later parts.}

The 8D transport $\capf$, or its variants, are sometimes called a BSSRDF (bidirectional subsurface scattering reflectance distribution function) in the literature. We will avoid this term, as it unnecessarily constrains the concept to subsurface scattering, while our solution generalizes to any other kinds of scattering.
%: glossy interreflection, fiber scattering, granular media, etc. We will simply refer to $\capf$ as the 8D light transport function.

\paragraph{Global light transport with a far-field assumption} 
assumes incident light is sufficiently distant to depend only on $\omegai$.
%Far-field approximations assume incident light $\capl_i(\omegai)$ sufficiently distant that it can be assumed to depend only on $\omegai$.
In this case,
$\capf$ can be further pre-integrated into a 6D transport $\fnear(\xxo,\omegao,\omegai)$, invariant to the incident position:
\begin{equation}
\begin{split}
%    \capl_o(\xxo,\omegao)&=\int_{\sphere} \int_{\surface} \capf(\xxo,\omegao,\xxi,\omegai)\capl_i(\omegai)\mathrm{d}\xxi\mathrm{d}\omegai\\
%    &=\int_{\sphere} \left(\int_{\surface} \capf(\xxo,\omegao,\xxi,\omegai)\mathrm{d}\xxi\right)\capl_i(\omegai)\mathrm{d}\omegai\\
%    &=\int_{\sphere} \fnear(\xxo,\omegao,\omegai)\capl_i(\omegai)\mathrm{d}\omegai.
    \capl_o(\xxo,\omegao;\capl_i)
    &=\iint \capf(\xxo,\omegao,\xxi,\omegai)\capl_i(\omegai)\mathrm{d}\xxi\mathrm{d}\omegai\\
    &=\int \left(\int \capf(\xxo,\omegao,\xxi,\omegai)\mathrm{d}\xxi\right)\capl_i(\omegai)\mathrm{d}\omegai\\
    &=\int\!\fnear(\xxo,\omegao,\omegai)\capl_i(\omegai)\mathrm{d}\omegai.
\end{split}
\label{eq:2-farfield}
\end{equation}%
Many far-field neural representations of this kind have been proposed in previous work. On flat materials, \citet{kuznetsov2021neumip} learn a spatially varying reflectance with a neural network $\fnear_\theta$ by minimizing an L2 regression loss between $\fnear_\theta$ and $\fnear$: 
\begin{equation}
\begin{split}
%\bm{\mathcal{L}}_{\capf}(\bm{\theta})=
    \iiint (\fnear_\theta(\xxo,\omegao,\omegai)
    -\fnear(\xxo,\omegao,\omegai))^2\mathrm{d}\omegai\mathrm{d}\xxo\mathrm{d}\omegao,
\end{split}
\label{eq:2-regression}
\end{equation}%
where the integral is estimated by sampling random $(\xxo,\omegao,\omegai)$ queries,
and the ground truth $\fnear$ is obtained via standard path tracing under a directional delta light source $\capl_i(\bo)=\delta(\bo-\omegai)$ (using the fact $\fnear(\xxo,\omegao,\omegai)\!=\!\int\!\fnear(\xxo,\omegao,\bo)\delta(\bo\!-\!\omegai)\mathrm{d}\bo$).
Similar ideas have been applied to layered materials~\cite{zeltner2024real,sztrajman2021neural,fan2022neural}, curved surfaces~\cite{kuznetsov2022rendering}, 3D assets with interreflection and scattering effects~\cite{mullia2024rna},
and to full-sphere incident directions for translucent objects~\cite{tg2024neupress}.
Note, this paper uses near-field specifically to refer to multi-bounce light transport rather than local BSDF models.

While the far-field assumption leads to a simple and effective learning step,
the resulting 6D model cannot accurately capture the near-field illumination effects encoded in the full transport $\capf$ of \cref{eq:2-glte}.
A direct extension would be to train an 8D network $\capf_\theta$ with the same regression scheme, now additionally sampling $\xxi$ and path tracing ground truth under a doubly delta illumination $\capl_i(\bx,\bo)=\delta(\bx-\xxi)\delta(\bo-\omegai)$,
but the combination of higher-dimensional sampling and strongly peaked illumination dramatically increases the difficulty of designing efficient Monte Carlo estimators.
While \citet{tg2024neural} partially mitigates this variance by learning a light transport function over a collection of incident illumination samples,
it relies on an isotropic assumption for internal scattering (see supplementary),
and its rendering requires tracing multiple incident rays per outgoing ray
that does not easily fit into the path tracing pipeline.
These limitations motivate our shift to a distribution-learning-based training objective.

\paragraph{Distribution learning in rendering}
is most commonly used for importance sampling.
These methods learn to sample a certain term of the rendering equation,
including BSDF sampling~\cite{rainer2020unified,xie2019multiple,xu2023neusample},
product sampling the BSDF times emission~\cite{herholz2016product,hart2020practical,clarberg2008practical,litalien2024neural},
and path guiding from the approximated incident radiance~\cite{vorba2014line,muller2017practical,dodik2022path,diolatzis2020practical}.
The distributions are constructed from analytic mixtures or warps in classic approaches.
In neural methods,
normalizing flows~\cite{dinh2014nice,rezende2015variational,durkan2019neural,muller2019neural} are widely used, yet still predict analytic mixtures for their efficiency\cite{sztrajman2021neural,xu2023neusample,fan2022neural,zeltner2024real}.
Recent works further explore flow-based models~\cite{lipman2022flow,heitz2023iterative} for BSDF sampling~\cite{fu2024importance} and precomputing multiple scattering~\cite{li2025puresample}. 
A reparameterization based neural sampling strategy is proposed in \cite{wu2025neural} without using generative neural networks.

%guiding~\cite{vorba2014line,muller2017practical,dodik2022path,diolatzis2020practical} that samples the approximated incident radiance rather than the emitter,
%product sampling the emitter times the BRDF term~\cite{herholz2016product,hart2020practical,clarberg2008practical}
%\lw{May be delete and cite additional works using analytic warps for distribution learning.}
%Beyond Monte Carlo sampling, generative models have also been applied to neural material generation~\cite{raghavan2025generative}, inverse rendering~\cite{zeng2024rgb}, and forward rendering~\cite{liang2025diffusion}.

\paragraph{Pre-computed radiance transfer (PRT)}
also builds upon the far-field light transport formulation of \cref{eq:2-farfield} but focuses on aggressively compressing both $L_i$ and $\mathcal{F}$~\cite{ramamoorthi2009precomputation}.
This is typically achieved by basis function expansions using spherical harmonics~\cite{sloan2023precomputed,ramamoorthi2001efficient}, wavelets~\cite{ng2003all,ng2004triple}, polynomials~\cite{ben2008precomputed}, Gaussians~\cite{tsai2006all,xu2013anisotropic,lu2025gaussian}, or neural bases~\cite{xu2022lightweight,rainer2022neural}.
Extensions to near-field polygonal area lights have also been derived using spherical harmonics~\cite{wang2018analytic,wu2020analytic}.
In contrast to path tracing, PRT is primarily designed for deterministic, real-time evaluation under relatively simple illumination configurations,
rather than for the high-fidelity, scalable simulation of complex light transport that we target.

\paragraph{Multiple-scattering by BSSRDF.}
BSSRDFs have been widely used to model specific multiple-scattering effects.
\citet{wann2001practical} first introduce an analytic BSSRDF for subsurface scattering in homogeneous media under flat surfaces.
Subsequent works refine this formulation~\cite{habel2013photon,donner2006spectral,donner2009empirical},
extend it to heterogeneous media~\cite{donner2008layered,peers2006compact},
and employ neural representations to better capture shape-dependent behavior~\cite{vicini2019learned}.
Similar BSSRDFs have also been used to model scattering inside translucent objects~\cite{deng2022reconstructing} and fur and hair~\cite{yan2017bssrdf}.
These methods typically adopt simplified BSSRDF forms (\eg separable parameterizations) and focus on a single class of effects (\eg subsurface scattering),
leaving other components of the light transport (\eg surface interreflections) to be handled by other methods or ignored.
In contrast, our 8D light transport makes no simplifications and is designed to capture all scattering events within the asset---across both interior volumes and surfaces--up to the point where light exits the asset.
\section{8D Neural Asset}
\label{sec:method}
Because the light transport function $\capf$ is non-negative,
we can decompose it to a conditional distribution $\mathbf{p}(\xxi,\omegai|\xxo,\omegao)$ and a normalizing factor $\bm{\alpha}(\xxo,\omegao)$ for each color channel independently:
%then learn them separately:
\begin{equation}
\begin{gathered}
\capf(\xxo,\omegao,\xxi,\omegai)=
\bm{\alpha}(\xxo,\omegao) \mathbf{p}(\xxi,\omegai | \xxo,\omegao),\\
\bm{\alpha}(\xxo,\omegao)\!=\!\!\iint\!\! \capf(\xxo,\omegao,\xxi,\omegai)\mathrm{d}\xxi\mathrm{d}\omegai,\; \mathbf{p}(\xxi,\omegai | \xxo,\omegao)=\frac{\capf}{\bm{\alpha}}.
\end{gathered}
\end{equation}%
Conceptually, $\bm{\alpha}$ corresponds to the directionally-varying albedo (survival probability) and $\mathbf{p}$ describes the distribution of scattering events.
%This decomposition is valid because $\capf$ is non-negative.
We learn the two components separately:
a normalizing flow $\mathbf{p}_{\bm{\theta}}$ for the scattering distribution and a regular MLP $\bm{\alpha}_{\bm{\theta}}$ for the albedo (\cref{subsec:3-learning}),
which together formulate $\capf_{\bm{\theta}}=\bm{\alpha}_{\bm{\theta}}\mathbf{p}_{\bm{\theta}}$.
Both $\xxi$ defined on the asset boundary and $\omegai$ on the sphere do not naturally fit into standard Euclidean normalizing flows.
We therefore apply a reparameterization of $\mathbf{p}_{\bm{\theta}}$ in \cref{subsec:3-generative}.
The training and inference of our model are described in \cref{subsec:3-architecture},
and we include their pseudocode in the supplemental material.

\begin{figure}[t]
    \centering
    \includegraphics[width=0.95\linewidth]{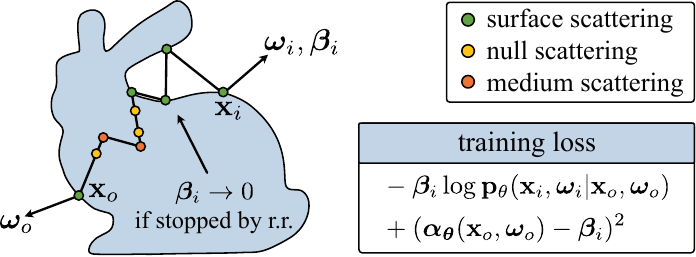}
    \caption{\textbf{Our training} is performed by tracing random outgoing rays to the asset, encountering multiple events of surface/medium/null scattering and Russian Roulette (r.r.) until leaving the asset.
    The exit ray configurations and throughputs are used to compute the negative log-likelihood loss and albedo regression loss (bottom right). 
    }
    \label{fig:3-train}
\end{figure}
\subsection{Learning the scattering distribution and albedo}
\label{subsec:3-learning}
We learn both $\mathbf{p}_{\bm{\theta}}$ and $\bm{\alpha}_{\bm{\theta}}$ through stochastic gradient descent optimization.
We first discuss the training objectives under a fixed $(\xxo,\omegao)$ then the combined loss for the full 8D domain.

\paragraph{Scattering distribution loss.}
The common practice in statistical learning suggests $\mathbf{p}_{\bm{\theta}}$ can be learned by minimizing the negative log-likelihood:
$-\!\!\iint\!\mathbf{p}\log\mathbf{p}_{\bm{\theta}}
\mathrm{d}\xxi\mathrm{d}\omegai
=-\frac{1}{\bm{\alpha}}\!\!\iint\!\capf\log\mathbf{p}_{\bm{\theta}}
\mathrm{d}\xxi\mathrm{d}\omegai$.
We drop $\tfrac{1}{\bm{\alpha}}$ as it is only a constant factor that does not affect the gradient-descent direction, which yields the loss:
%Dropping the constant factor $\tfrac{1}{\bm{\alpha}}$ yields the loss:
\begin{equation}
\small
\bm{\mathcal{L}}_\mathbf{p}(\bm{\theta}
|\xxo,\omegao
)\!= -\!\!\iint\!\!\capf(\xxi,\omegai,\xxo,\omegao)
    \log \mathbf{p}_{\bm{\theta}} (\xxi,\omegai | \xxo,\omegao)\mathrm{d}\xxi\mathrm{d}\mathbf{\omega}_i.
\label{eq:3-distribution}
\end{equation}
Comparing the right-hand side of \cref{eq:3-distribution} with \cref{eq:2-glte},
we can interpret $\bm{\mathcal{L}}_\mathbf{p}$ as rendering the asset with $-\log \mathbf{p}_{\bm{\theta}}$ acting as the "incident light" (can be positive or negative):
%\begin{align}
%\iint\!\!&\capf(\xxo,\omegao,\xxi,\omegai)
%(\underbrace{-\log\mathbf{p}_{\bm{\theta}}(\xxi,\omegai|\xxo,\omegao)}_{\capl_i})\mathrm{d}\xxi\mathrm{d}\mathbf{\omega}_i\tag*{}\\[-3ex]
%&=\capl_{pt}(\xxo,\omegao,-\log\mathbf{p}_{\bm{\theta}})
%\tag*{}
%\end{align}
\begin{align}
-\!\iint\!\!\capf
\log\mathbf{p}_{\bm{\theta}}\mathrm{d}\xxi\mathrm{d}\mathbf{\omega}_i\!=
\!\!\iint\!\!\capf(-\log\mathbf{p}_{\bm{\theta}})
%({-\log\mathbf{p}_{\bm{\theta}}})
\mathrm{d}\xxi\mathrm{d}\mathbf{\omega}_i=\capl_{o}(\xxo,\omegao;-\log\mathbf{p}_{\bm{\theta}}).
\tag*{}
\end{align}%
Using the fact that \cref{eq:2-glte} can be unbiasedly estimated through path tracing,
we can compute $\bm{\mathcal{L}}_\mathbf{p}$ in a similar path-tracing style (\cref{fig:3-train}):
the ray $(\xxo,\omegao)$ is traced (path-sampled) for multiple bounces until leaving the asset,
whose exit configuration $(\xxi,\omegai)$ and throughput $\bm{\beta}_i$ is then used to form the Monte Carlo estimator ($\mathbb{E}^n$ denotes empirical expectation over $n$ samples):
\begin{equation}
   \bm{\mathcal{L}}_\mathbf{p}(\bm{\theta}
   |\xxo,\omegao)=\!\!\mathop{\mathbb{E}^n}_{\xxi,\omegai,\bm{\beta}_i\sim \text{path sampling}}
    \!\!\left[
    -\bm{\beta}_i \log \mathbf{p}_\theta(\xxi,\omegai|\xxo,\omegao)
    \right].
\label{eq:3-distribution-MC}
\end{equation}%
The path-sampling probability does not need to equal $\mathbf{p}$ since $\bm{\beta}_i$ gives the correction,
and the throughput-based Russian roulette also seamlessly fits here to terminate low-throughput paths early,
for which the terminated paths simply contribute $\bm{\beta}_i=0$.

\paragraph{Albedo loss.}
Under the same intuition, the albedo can be interpreted as rendering the asset under constant unit illumination:
\begin{equation}
\bm{\alpha}(\xxo,\omegao)=\iint\!\capf(\xxo,\omegao,\xxi,\omegai)\times\mathbf{1}\mathrm{d}\xxi\mathrm{d}\omegai=\capl_{o}(\xxo,\omegao;\mathbf{1})
\tag*{}
\end{equation}
which can be estimated by averaging the throughput $\bm{\alpha_\theta}(\xxo,\omegao)=\mathbb{E}^n[\bm{\beta}_i]$,
reusing the same samples for training $\mathbf{p}_{\bm{\theta}}$.
%We therefore train $\bm{\alpha}_{\bm{\theta}}$ with a simple L2 regression loss:
We fit $\bm{\alpha}_{\bm{\theta}}$ against the estimated $\bm{\alpha}$ using a simple L2 loss:
\begin{equation}
\begin{split}
    \bm{\mathcal{L}}_{\bm{\alpha}}(\bm{\theta}|\xxo,\omegao)&=(\bm{\alpha}_{\bm{\theta}}(\xxo,\omegao)-\bm{\alpha}(\xxo,\omegao))^2\\
    &=(\bm{\alpha_\theta}(\xxo,\omegao)-\mathbb{E}^n[\bm{\beta}_i])^2,
\end{split}
\label{eq:3-albedo-loss}
\end{equation}%
whose gradient is linear in $\bm{\alpha}$,
therefore the empirical expectation gives an unbiased gradient estimator.
%so can be unbiasedly estimated using the empirical expectation.

\begin{figure}[t]
    \centering
    \setlength\tabcolsep{0.5pt}
    \resizebox{0.99\linewidth}{!}{
    \begin{tabular}{cccc}
        Reference & Ours & Regression & Regression\\
        (training sample) & (1 sample) & (8192 sample) & (1 sample)\\
         \includegraphics[width=0.3\linewidth]{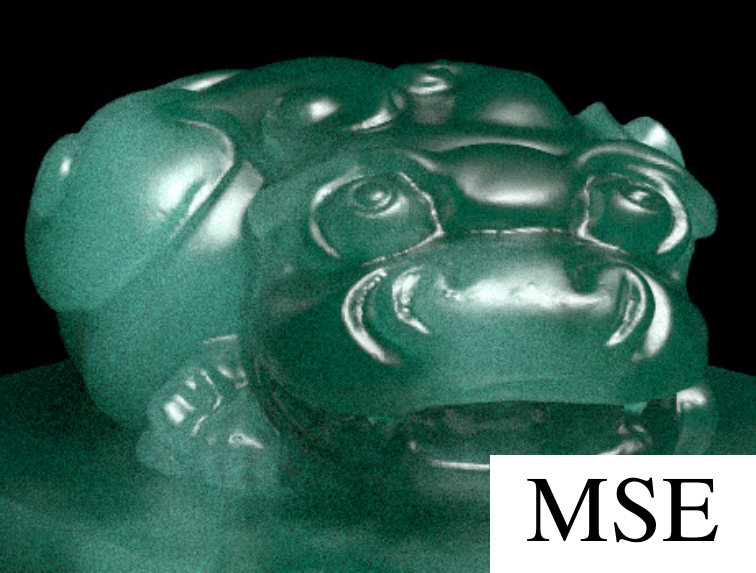}&
         \includegraphics[width=0.3\linewidth]{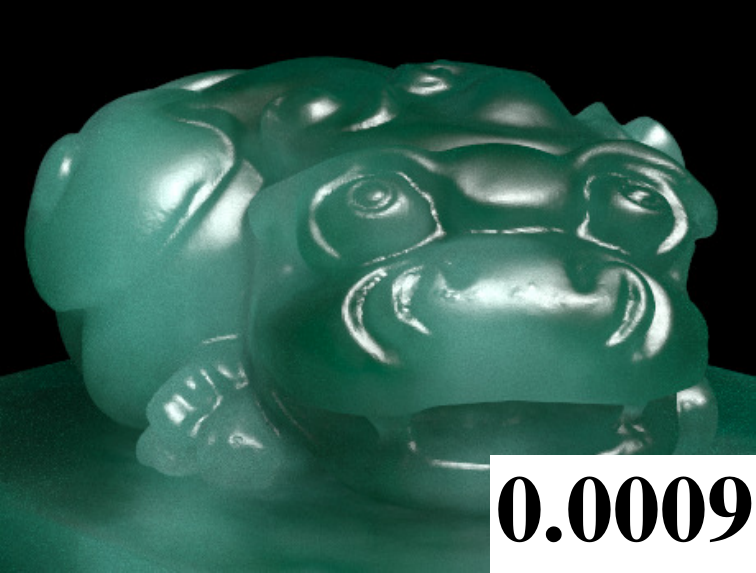}&
         \includegraphics[width=0.3\linewidth]{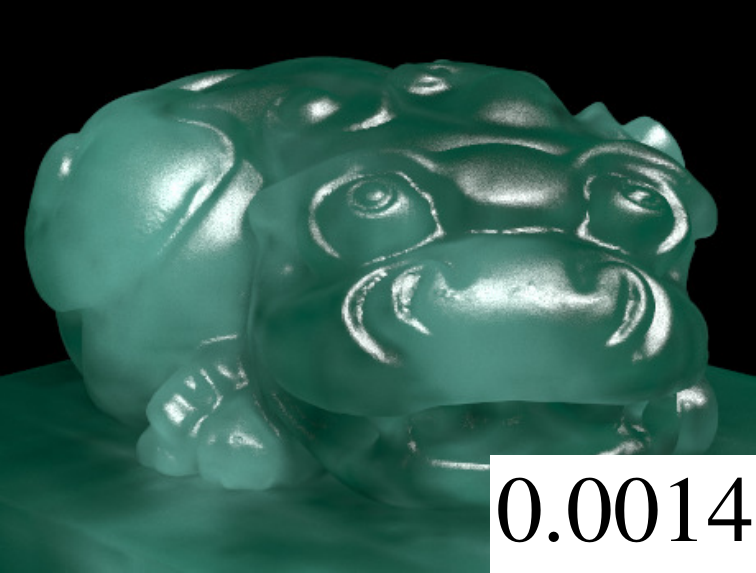}&
         \includegraphics[width=0.3\linewidth]{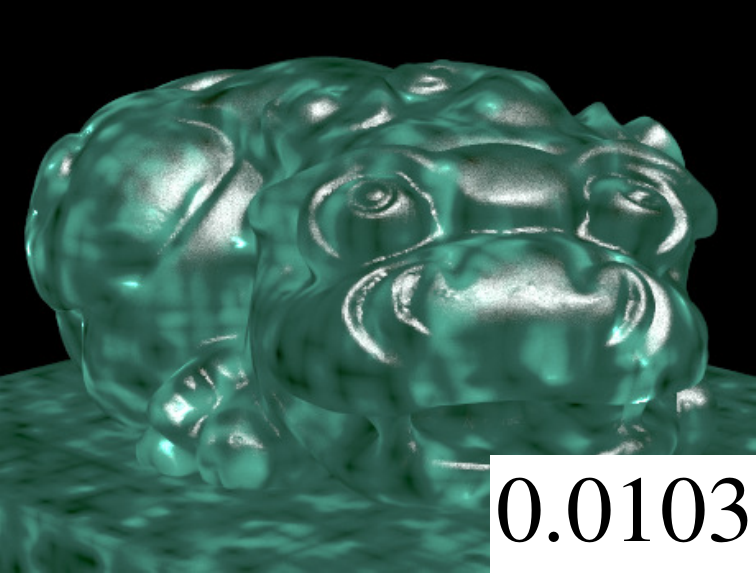}\\
    \end{tabular}
    }
    \caption{
    \textbf{Our training vs. regression.}
    Regressing the ground truth light transport requires many samples per training query to estimate $\fnear$ (3rd image),  which fails in the 1-sample setup (4th image). 
    In contrast, our optimization requires only 1 sample per $(\xxo,\omegao)$ query (2nd image).
    The numbers show the mean square error (MSE), and a far-field light is used.
    }
    \label{fig:3-train-spp}
\end{figure}
\begin{figure}[t]
    \centering
    \includegraphics[width=0.99\linewidth]{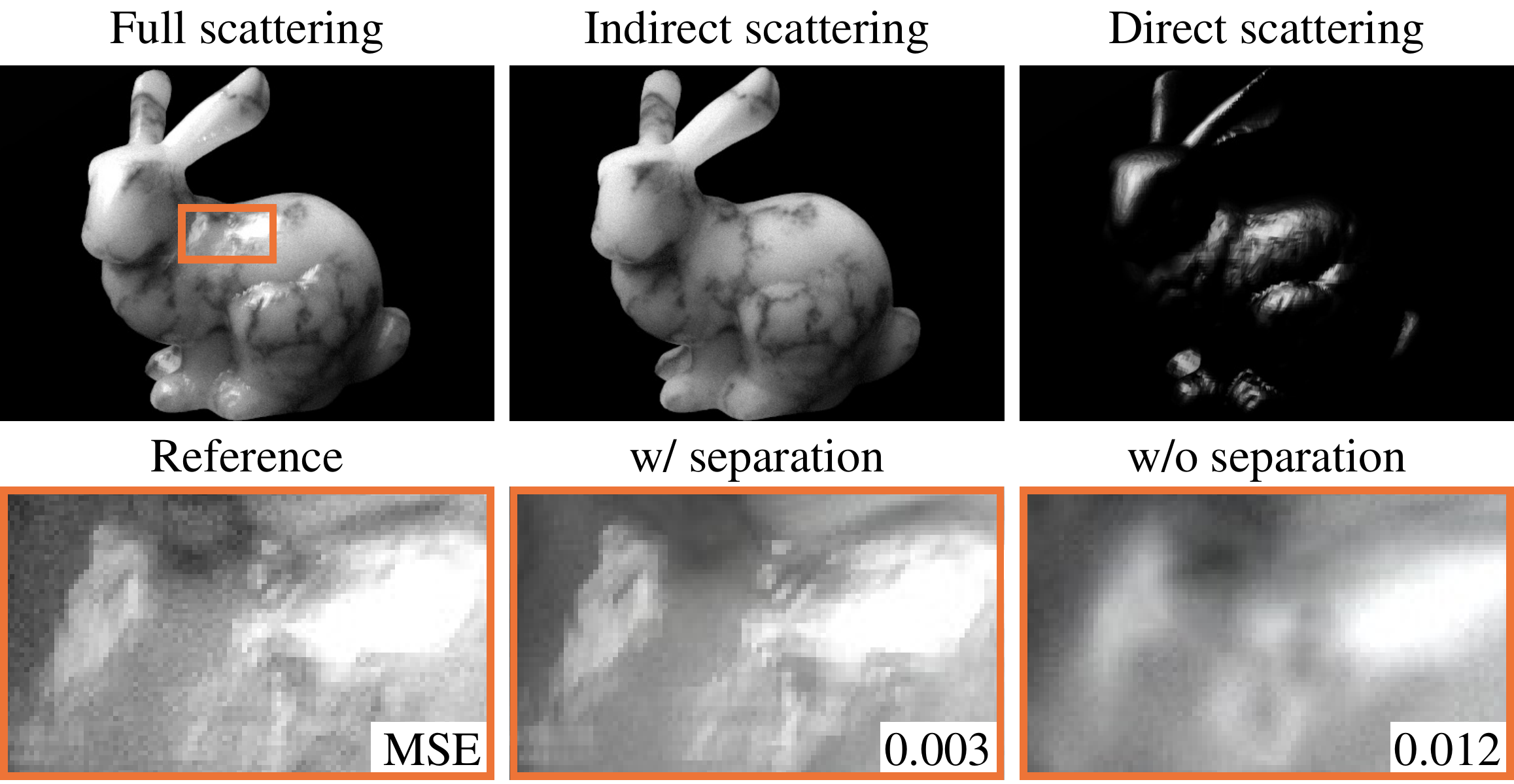}
    \caption{\textbf{Direct-indirect separation of light transport.}
    The direct scattering above is nearly a delta reflection, while the indirect light transport involves smooth volumetric scattering.
    Such different behaviors are difficult to model by a single network (w/o separation), so we only model the indirect component combined with analytic direct scattering to improve the accuracy (w/ separation).
    More direct/indirect examples are in the supplement.
    }
    \label{fig:3-separation}
\end{figure} 
\paragraph{Loss variance analysis.}
Figure~\ref{fig:3-train-spp} shows an example where the regression objective \cref{eq:2-regression} requires thousands of samples to estimate $\fnear$ for convergence,
whereas our model trained with one sample per outgoing ray already recovers $\capf$.
This is because the regression gradient $2\!\int\!(\fnear_{\bm{\theta}}-\fnear)\nabla_{\bm{\theta}}\fnear_{\bm{\theta}}\mathrm{d}\omegai$ lacks an efficient way to both sample $\omegai$ and estimate $\fnear$ for arbitrary $\omegai$,
making the optimization especially hard for highly specular light transport induced by dielectric boundaries, fibers, \etc
In contrast, our gradient $\nabla_{\bm{\theta}}\bm{\mathcal{L}}_{\mathbf{p}}\!\propto\!\iint\!\capf\nabla_{\theta}\log\mathbf{p}_{\bm{\theta}}\mathrm{d}\xxi\mathrm{d}\omegai$ and $\nabla_{\bm{\theta}}\bm{\mathcal{L}}_{\bm{\alpha}}\!\propto\!(\bm{\alpha}_{\bm{\theta}}-\!\iint\!\capf\mathrm{d}\xxi\mathrm{d}\omegai)\nabla_{\bm{\theta}}\bm{\alpha}_{\bm{\theta}}$ are simple integrals against $\capf$.
This is what is importance-sampled by the path-sampling procedure,
therefore \cref{eq:3-albedo-loss,eq:3-distribution-MC} admit empirically low-variance gradient estimators suitable for stochastic optimization.
This sampling efficiency ensures our model is trainable under the 8D configuration.

\paragraph{Overall loss.}
To optimize $\mathbf{p}_{\bm{\theta}},\bm{\alpha}$ jointly across all outgoing configurations,
we integrate $\bm{\mathcal{L}}_{\mathbf{p}}+\bm{\mathcal{L}}_{\bm{\alpha}}$ over $\xxo,\omegao$,
estimated by sampling random outgoing rays that intersect the asset (\cref{subsec:3-architecture}),
and sum the contributions over color channels $c$ (superscripts denote $c$-th component of vectors):
%\begin{equation}
%\small
%    \mathcal{L}(\bm{\theta})=\!\!\sum_{\forall c}
%    \iint \left(\bm{\mathcal{L}}_{\mathbf{p}}^c(\bm{\theta}|\xxo,\omegao)\!+\!\bm{\mathcal{L}}_{\bm{\alpha}}^c(\bm{\theta}|\xxo,\omegao)\right)\mathrm{d}\xxo\mathrm{d}\omegao.
%\label{eq:3-full-loss}
%\end{equation}

\begin{equation}
    \mathcal{L}(\bm{\theta})=\!\!\sum_{\forall c}
     \mathop{\mathbb{E}^n}_{\xxo,\omegao}\left[
\bm{\mathcal{L}}_{\mathbf{p}}^c(\bm{\theta}|\xxo,\omegao)\!+\!\bm{\mathcal{L}}_{\bm{\alpha}}^c(\bm{\theta}|\xxo,\omegao)\right].
\label{eq:3-full-loss}
\end{equation}%

\paragraph{Separating the direct scattering.}
Translucent objects often have simple but highly specular dielectric boundaries whose direct scattering is much sharper than the rest of the light transport (\cref{fig:3-separation}). 
To better capture the remaining complexity,
we keep the (cosine-weighted) direct BSDF $\mathbf{f}$ untouched and train the network only on the indirect transport $\capf(\xxo,\omegao,\xxi,\omegai)\!-\!V(\xxo,\omegai)\mathbf{f}(\xxo,\omegao,\omegai)$,
where $V$ denotes the asset's self-visibility.
This is implemented by using the same loss \cref{eq:3-full-loss} while setting $\bm{\beta}_i=0$ for path samples that exit the asset after one scattering event.
At inference time, $V\mathbf{f}$ is still evaluated analytically, which is usually cheap, 
yet this direct-indirect separation leads to more accurate matching of the ground truth appearance.

\subsection{Realizing the distribution network}
\label{subsec:3-generative}
\begin{figure}[t]
    \centering
    \includegraphics[width=0.95\linewidth]{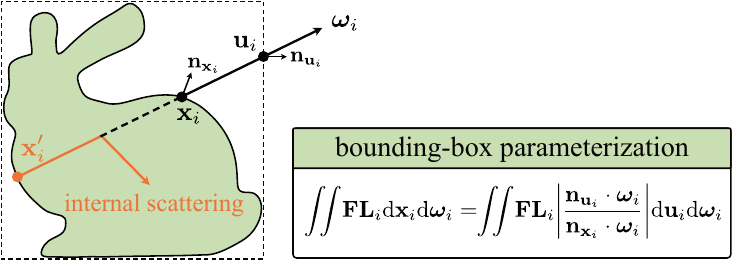}
    \caption{\textbf{Bounding-box parameterization of $\xxi$.}
    We map $\xxi$ along $\omegai$ to its bounding-box intersection  $\mathbf{u}_i$ and reparameterize \cref{eq:2-glte} accordingly.
    While tracing the ray back from $\mathbf{u}_i$ may hit multiple points (\eg $\xxi,\bx'_i$),
    only the outermost intersection ($\xxi$) is visible to the incident light and lies in the original integral domain;
    other points contribute only to internal scattering.
    Thus, the mapping from $(\xxi,\omegai)$ to $(\mathbf{u}_i,\omegai)$ is injective, assuming incident illumination originates outside the asset convex hull (\cref{subsec:4-limitation}).
    }
    \label{fig:3-bbox}
\end{figure}

\paragraph{Parameterization of network inputs.}
%Both $\xxi$ (on the asset surface) and $\omegai$ lie on manifolds that need to be reparameterized for standard Euclidean normalizing flows.
%To that end, 
To obtain a Euclidean parameterization of $(\xxi,\omegai)$,
we trace the incident ray further to its intersection $\mathbf{u}_i$ on the asset's axis-aligned bounding box (assume centered at origin),
inducing the transformed pdf involving the surface normal $\mathbf{n}_\square$ (the subscript indicates the query location):
\begin{equation}
\begin{split}
\mathbf{p}_{\bm{\theta}}(
\xxi,\omegai|
\xxo,\omegao
)
%\mathrm{d}\xxi\mathrm{d}\omegai
\!=\!
\mathbf{p}_{\bm{\theta}}(
\mathbf{u}_i,\omegai|
\xxo,\omegao
)
\left\vert\frac{
\mathbf{n}_{\xxi}\cdot\omegai
}{
\mathbf{n}_{\mathbf{u}_i}\cdot\omegai
}\right\vert.
%\mathrm{d}\mathbf{u}_i\mathrm{d}\omegai.
\end{split}
\label{eq:3-xo-parameterization}
\end{equation}%
This is a valid one-to-one reparameterization as explained in \cref{fig:3-bbox};
other proxies such as a bounding sphere could be used similarly.
The proxy uv-map gives an general Euclidean parameterization of the intersection,
but for a bounding box,
we can simply encode $\mathbf{u}_i$ and $\omegai$ in the asset's local space to their cylindrical coordinates: 
%We then encode the pair $(\mathbf{u}_i,\omegai)$ in cylindrical coordinates $\mathbf{s}$, yielding a final Euclidean parameterization for learning:
\begin{equation}
\begin{gathered}
\mathbf{p}_{\bm{\theta}}(
\mathbf{u}_i,\omegai|
\xxo,\omegao
)
=
\mathbf{p}_{\bm{\theta}}(\mathbf{s}|
\xxo,\omegao
)
\frac{\vert\mathbf{n}_{\mathbf{u}_i}\cdot \mathbf{u}_i\vert}
{(\sqrt{\mathbf{u}_i\cdot\mathbf{u}_i})^3},
\\
\mathbf{s}=
\left(\mathbf{u}_i^3/\sqrt{\mathbf{u}_i\cdot\mathbf{u}_i},\arctan(\mathbf{u}_i^1/\mathbf{u}_i^2),\omegai^3,\arctan(\omegai^1/\omegai^2)\right).
\end{gathered}
\label{eq:3-cylindrical}
\end{equation}%
We do not use spherical coordinates as they have zero Jacobian determinants at the poles which introduces singularities.
Derivations of the projection factors above are described in the supplement.

%Other parameterizations like the geometry proxy uv-map can also be used as long as it is Euclidean.

\begin{figure}[t]
    \centering
    \includegraphics[width=0.95\linewidth]{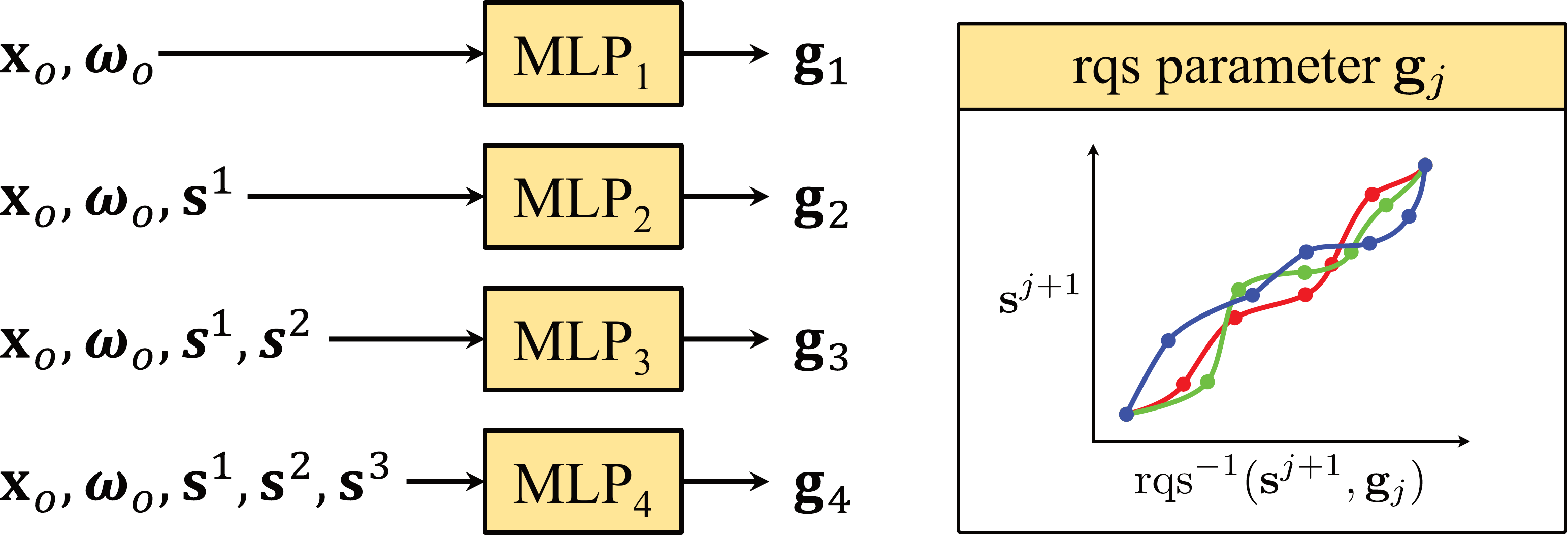}
    \caption{\textbf{Vector-valued normalizing flow for $\mathbf{p}_{\bm{\theta}}$}.
    Left: the distribution of $\mathbf{s}$ is constructed autoregressively using rational quadratic splines (rqs).
    Right: the rqs parameter $\mathbf{g}_j$ contains K knots for each RGB channel.
    }
    \label{fig:3-vnf}
\end{figure}

\paragraph{Vector-valued normalizing flow.}
Given the Euclidean parameter $\mathbf{s}$,
we model the scattering distribution with an autoregressive normalizing flow 
$\mathbf{p}_{\bm{\theta}}(
\mathbf{s}|\xxo,\omegao
)
=\prod_{j=1}^{4}
\mathbf{p}_{\bm{\theta}}(
\mathbf{s}^j| \mathbf{s}^{k<j}, \xxo,\omegao
)$.
As demonstrated in \cref{fig:3-vnf},
each conditional factor is realized by an MLP-predicted rational quadratic spline (rqs)~\cite{durkan2019neural} of knots $\mathbf{g}_j=\text{MLP}_{j,\bm{\theta}}(
\mathbf{s}^{k<j},\xxo,\omegao)$,
used for both pdf evaluation and sampling per color channel $c$:
\begin{equation}
\begin{split}
\quad \mathbf{p}^c_{\bm{\theta}}
(\mathbf{s}^j|\mathbf{s}^{k<j},\xxo,\omegao)
=\frac{\mathrm{d}\text{rqs}^{-1}(\mathbf{s}^j,\mathbf{g}_j^c)}{\mathrm{d}\mathbf{s}^j}\quad  & \text{(pdf evaluation)}\\
\mathbf{s}^j=\text{rqs}(u,\mathbf{g}_j^c) \quad u\sim \text{Uniform}(0,1)\quad  & \text{(pdf sampling)}.
\end{split}
\label{eq:3-rqs}
\end{equation}%
To sample $\mathbf{s}$, we draw $\mathbf{s}^1$ then $\mathbf{s}^2$ conditioned on $\mathbf{s}^1$ and so on,
and the full pdf evaluation is the product of the four derivatives .
Unlike standard normalizing flows that output a single scalar density,
our $\mathbf{p}_{\bm{\theta}}$ is vector-valued over RGB, so each $\mathbf{g}_j$ encodes three splines accordingly.
The autoregressive structure further implies a factorization of the distribution $\mathbf{p}_{\bm{\theta}}(\mathbf{u}_i,\omegai|\cdot)=\mathbf{p}_{\bm{\theta}}(\mathbf{u}_i|\cdot)\mathbf{p}_{\bm{\theta}}(\omegai|\mathbf{u}_i,\cdot)$, which we later exploit for multiple importance sampling (MIS) in \cref{subsec:3-architecture}.

\subsection{Training and inference}
\label{subsec:3-architecture}

\paragraph{Network architecture.}
Each of the 4 MLPs in the distribution network uses ReLU activations,
4 hidden layers, and 128 hidden features.
The $j$-th MLP takes $(\xxo,\omegao,\mathbf{s}^1,\cdots,\mathbf{s}^{j-1})$ as input and predicts the 2D knot positions and derivatives of a rational quadratic spline.
We use 32 spline knots per RGB channel, resulting in $3\times32\times(2+1)=288$ output size.
The albedo MLP takes $(\xxo,\omegao)$ as input and outputs an RGB color.
It also uses ReLU activations and 128 hidden features but with 2 hidden layers.

\paragraph{Input encoding.}
To capture high-frequency details,
$\xxo$ in all networks is encoded with a $3\times64\times64$ triplane feature grid~\cite{chan2022efficient},
and $\omegao$ is encoded with a $6\times32\times32$ cubemap feature grid~\cite{wu2024neural}.
In the distribution network,
$\mathbf{s}^1, \mathbf{s}^3$ in the first and third MLPs are encoded with a 1D feature grid of resolution 32.
For the third and fourth MLPs, the pair $(\mathbf{s}^1,\mathbf{s}^2)$ is first mapped to a unit direction then encoded with a $6\times32\times32$ cubemap grid.
All the encodings use feature vectors of size 8 per grid vertex.

\paragraph{Training data generation.}
We optimize \cref{eq:3-full-loss} using a buffer of path-sampled tuples $(\xxo,\omegao,\xxi,\omegai,\bm{\beta}_i)$ that are regenerated once being fully consumed.
Each tuple is obtained by first uniformly sampling a point on the asset's bounding sphere then cosine-weighted sampling $\omegao$ in the local tangent frame.
We trace this ray to find the first intersection $\xxo$ and continue path tracing to obtain $(\xxi,\omegai,\bm{\beta}_i)$.
Tuples whose initial ray misses the asset are excluded from the training.
Compared to precomputing the training data~\cite{mullia2024rna},
the online sampling prevents overfitting to a static set of paths.
The buffer stores $128^4$ such tuples in 15GB RAM.

\paragraph{Optimization.}
Our code is implemented using Pytorch~\cite{paszke2019pytorch} and Mitsuba 3~\cite{jakob2022dr}.
All networks are trained with the Adam optimizer~\cite{kingma2014adam} at a $0.0005$ learning rate for 240K steps, using a batch size of 32768 samples per step.
This leads to $(240\text{K}\times32768)/128^4\!=\!30$ buffer refreshes.

\begin{figure}[t]
    \centering
    \includegraphics[width=0.99\linewidth]{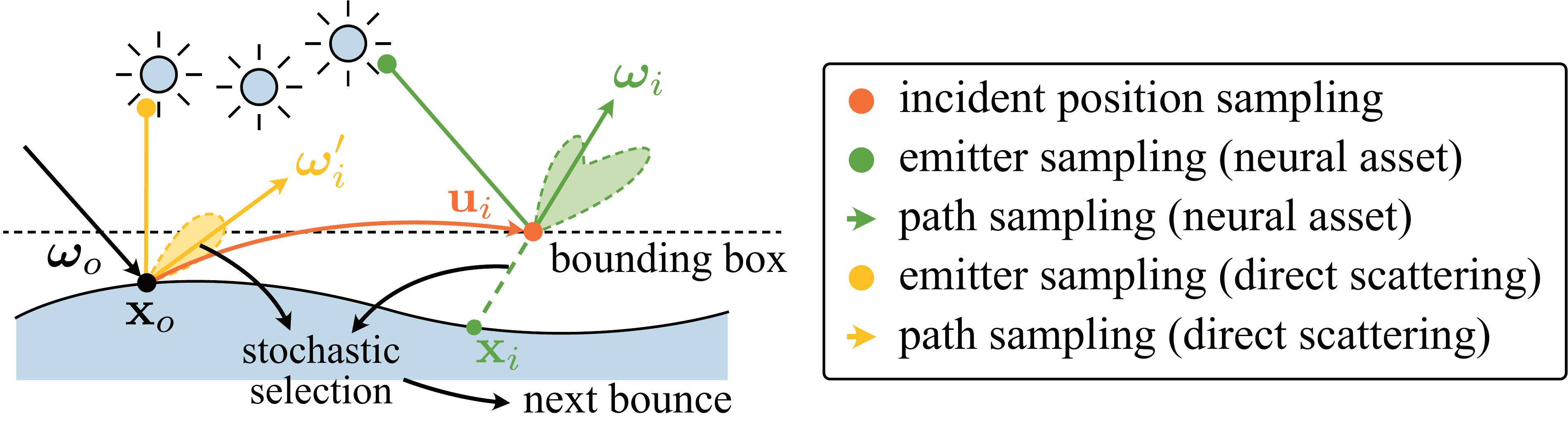}
    \caption{\textbf{Path tracing our model} is similar to the standard scheme, except we perform MIS at the sampled incident location (orange dot).
    If direct scattering is separated, we further emitter-sample the direct lobe at $\xxo$ (yellow dot) and stochastically choose between direct samples (yellow arrow) and neural asset samples (green arrow) in path sampling. 
    }
    \label{fig:3-pathtracing}
\end{figure}
\paragraph{Integrating with path tracing.}
Path tracing with our model follows a standard emitter-path sampling scheme (\cref{fig:3-pathtracing}):
each path-tracing bounce begins by sampling an incident location $\mathbf{u}_i\!\sim\!\mathbf{p}^c_{\bm{\theta}}(\mathbf{u}_i|\xxo,\omegao)$ using the pdf factorization in \cref{subsec:3-generative} over a uniformly sampled color channel $c$;
MIS is then performed at $\mathbf{u}_i$ between the emitter-sampling technique and $\mathbf{p}_{\bm{\theta}}^c(\omegai|\mathbf{u}_i,\xxo,\omegao)$.
When $\capf_\mathbf{\theta}$ is trained on the indirect scattering,
we also emitter-sample the direct lobe $\mathbf{f}$ at $\xxo$ and stochastically choose between a direct path sample and a neural asset sample as the next-bounce ray.
Details of our path tracing is provided in the supplemental material.
We use a custom Mitsuba integrator for path tracing and a CUDA kernel for the pdf sampling and evaluation (\cref{eq:3-rqs}).
We did not find a benefit of writing CUDA kernels for the MLPs given their relatively large sizes.
\section{Experiments}
\label{sec:experiments}
\begin{figure}[t]
    \centering
    \includegraphics[width=0.99\linewidth]{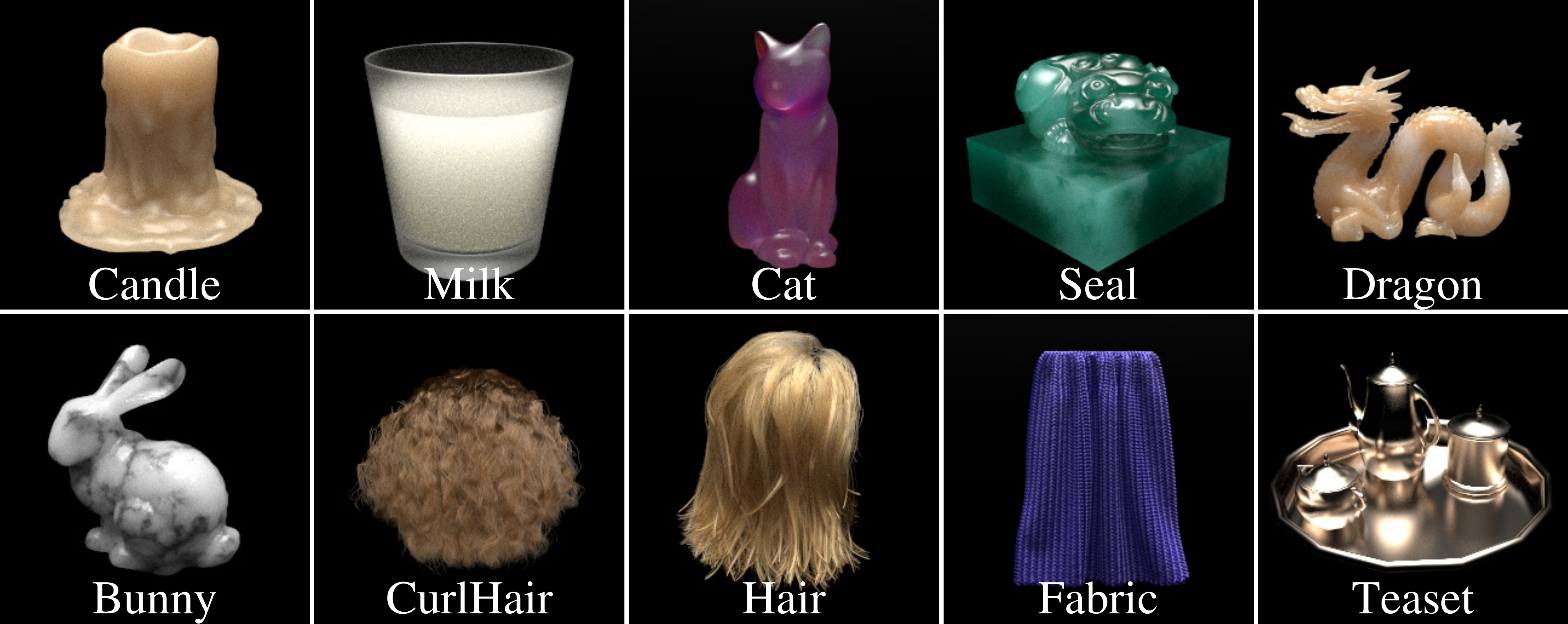}
    \caption{\textbf{Assets used in our experiments.} 
    From left to right, top to bottom, the first 6 assets exhibit strong subsurface scattering from homogeneous (\textit{Candle}, \textit{Milk}) and heterogeneous media (the rest), all enclosed by glossy dielectric boundaries.
    \textit{CurlHair}, \textit{Hair}, and \textit{Fabric} are modeled by a hair BSDF~\cite{chiang2016practical}, and \textit{Teaset} uses a conductor BSDF.
    }
    \label{fig:4-asset}
\end{figure}

For each asset in \cref{fig:4-asset},
we prebake its light transport to our model then evaluate its Monte Carlo rendering under novel scenes by comparing reconstruction accuracy and variance separately.
The reconstruction accuracy is measured against the non-prebaked path tracing reference in MSE at high spp to suppress variance.
The rendering variance is estimated by the MSE between the rendering and a high-spp rendering of the same method,
and it is considered together with rendering speed to evaluate sampling efficiency.
%which is reported together with rendering speed to demonstrate sampling efficiency.
We also measure the time required to prebake each asset.
All images are rendered in $800\times800$ resolution on an NVIDIA 3090 GPU.

\subsection{Baselines.}
We mainly compare with the standard path tracing (PT) and a far-field light transport baseline (\methodA{}) constructed following \citet{tg2024neupress} and \citet{mullia2024rna}.
While \cite{tg2024neural} is also a near-field method,
it is restricted to isotropic subsurface scattering and does not generalize to all of our test scenes. We therefore include only a partial comparison to this method in the supplement.

\paragraph{Far-field.}
We model the 6D $\fnear$ with an MLP of 128 hidden units and 8 hidden layers trained over the regression loss \cref{eq:2-regression};
and we learn the importance sampling of $\omegai$ following \citet{xu2023neusample},
using a 16-knot rqs normalizing flow parameterized by two MLPs (each with 64 hidden units and 2 hidden layers).
The overall network capacity is chosen to be comparable to ours excluding $\mathbf{p}_{\bm{\theta}}(\mathbf{u}_i|\xxo,\omegao)$ (as it is not presented in $\fnear$),
to ensure the performance gain is from the representation rather than over-parameterization.
All networks use the same input encodings and optimization settings as described in \cref{subsec:3-architecture}.
We do not use the visibility hint~\cite{muller2019neural} but apply our direct-scattering separation that serves the same purpose.
To obtain ground-truth $\fnear$ for supervision,
we find 4096 spp are required for convergence on purely surface-scattering assets and 8192 spp for assets with volumetric interiors.
Generating and resampling such data online would take several days~\cite{tg2024neupress},
so we precompute a static dataset of $1024\times256^2$ training samples using the strategy of \citeauthor{mullia2024rna}.

\paragraph{Path tracing.}
We disable emitter sampling for media enclosed in dielectric boundaries,
as the rays inside are never visible to light sources.
%For a fair speed comparison,
For fair comparison,
we also disable the Mitsuba megakernel backend when recording rendering speed that is not available for the other methods involving PyTorch calls.

\begin{table}[t]
    \centering
    \setlength\tabcolsep{2 pt}
    \caption{\textbf{Quantitative rendering error.} Our model achieves lower MSE (scaled by 100) on all assets than the far-field baseline.}
    \resizebox{0.99\linewidth}{!}{
    \begin{tabular}{l c c c c c c c c c c}
        \toprule
        \multirow{2}{*}{\textbf{Method}} 
        &
        Candle & Milk & Cat & Seal & Dragon & Bunny & CurlHair & Hair & Fabric & Teaset\\
        &\multicolumn{10}{c}{\textbf{100$\times$MSE$\downarrow$}}\\
        \midrule
\textbf{\methodA{}} & 5.760 & 0.430 & 0.162 & 0.065 & 0.505 & 0.103 & 0.228 & 1.381 & 3.379 & 0.707\\
\textbf{Ours} & 2.855 & 0.156 & 0.009 & 0.057 & 0.182 & 0.057 & 0.126 & 0.950 & 2.541 & 0.075\\
        \bottomrule
    \end{tabular}
    }
    \label{tab:4-accuracy}
\end{table}
\begin{table}[t]
    \centering
    \setlength\tabcolsep{2 pt}
    \caption{\textbf{Rendering variance and speed at 128 spp.}
    Far-field model achieves the lowest variance and fastest inference speed but suffers from biased light transport.
    Our rendering achieves accuracy comparable to path tracing while reducing variance and rendering time: roughly $2$-$20\times$ faster on volumetric assets and $1.4$-$4\times$ faster on fine-scale surface scattering (\textit{CurlHair}, \textit{Hair}, \textit{Fabric}).
    All methods render simple assets (\textit{Teaset}) efficiently.
    }
    \resizebox{0.99\linewidth}{!}{
    \begin{tabular}{l c c c c c c c c c c}
        \toprule
        \multirow{2}{*}{\textbf{Method}} 
        &
        Candle & Milk & Cat & Seal & Dragon & Bunny & CurlHair & Hair & Fabric & Teaset\\
        &\multicolumn{10}{c}{\textbf{100$\times$Variance$\downarrow$ (128spp)}}\\
        \midrule
\textbf{\methodA{}} & 12.20 & 0.445 & 0.093 & 0.011 & 0.134 & 0.014 & 0.134 & 0.067 & 0.090 & 0.110\\
\textbf{PT} & 54.34 & 33.41 & 0.115 & 2.561 & 5.723 & 3.673 & 0.655 & 2.219 & 0.414 & 0.102\\
\textbf{Ours} & 27.44 & 2.491 & 0.094 & 0.084 & 1.975 & 0.588 & 0.572 & 2.023 & 0.573 & 0.107\\
        \midrule
         &\multicolumn{10}{c}{\textbf{minutes$\downarrow$ (128spp)}}\\
        \midrule
\textbf{\methodA{}} & 0.35 & 0.39 & 0.33 & 0.30 & 0.35 & 0.35 & 0.29 & 0.30 & 0.40 & 0.40\\
\textbf{PT} & 4.13 & 15.6 & 1.28 & 2.04 & 4.27 & 4.97 & 0.80 & 2.12 & 4.49 & 0.82\\
\textbf{Ours} & 0.53 & 0.69 & 0.57 & 0.55 & 0.52 & 0.57 & 0.57 & 0.58 & 0.93 & 0.65\\
        \bottomrule
    \end{tabular}
    }
    \label{tab:4-efficiency}
\end{table}
\begin{figure*}[p]
    \centering
    \includegraphics[width=0.99\linewidth]{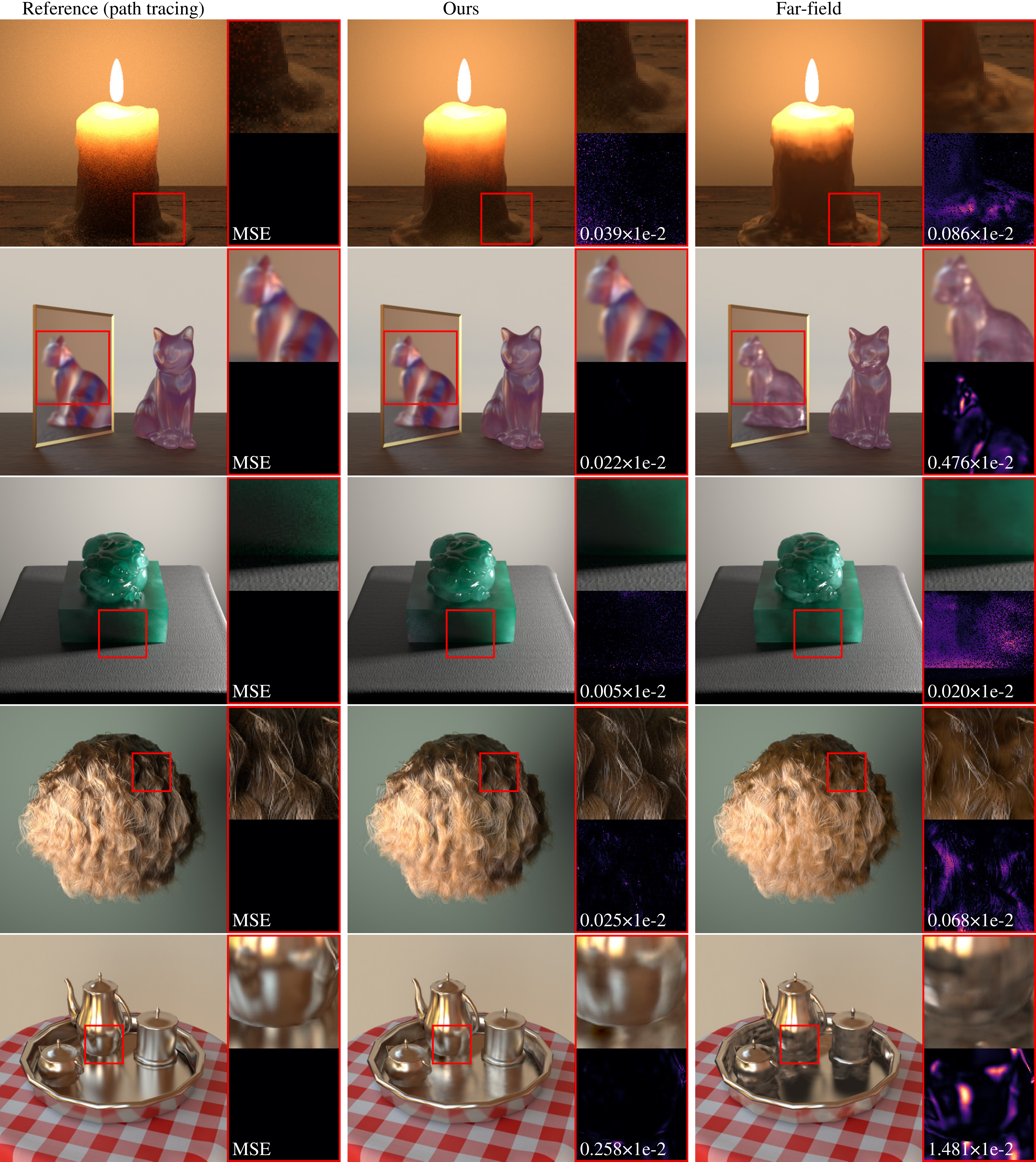}
    \caption{\textbf{Qualitative rendering comparison.} 
    The assets shown in \cref{fig:4-asset} are baked into neural light transport models, and the background geometry uses standard material models.
    We use 8192 spp except for \textit{Milk}, \textit{Seal}, and \textit{Dragon} where the path tracing uses 65536 spp.
    Our method more accurately captures the light transport of the assets under near-field illumination.
    The numbers show the MSEs of the insets.}
    \label{fig:4-acc}
\end{figure*}
\begin{figure*}[t]
    \includegraphics[width=0.99\linewidth]{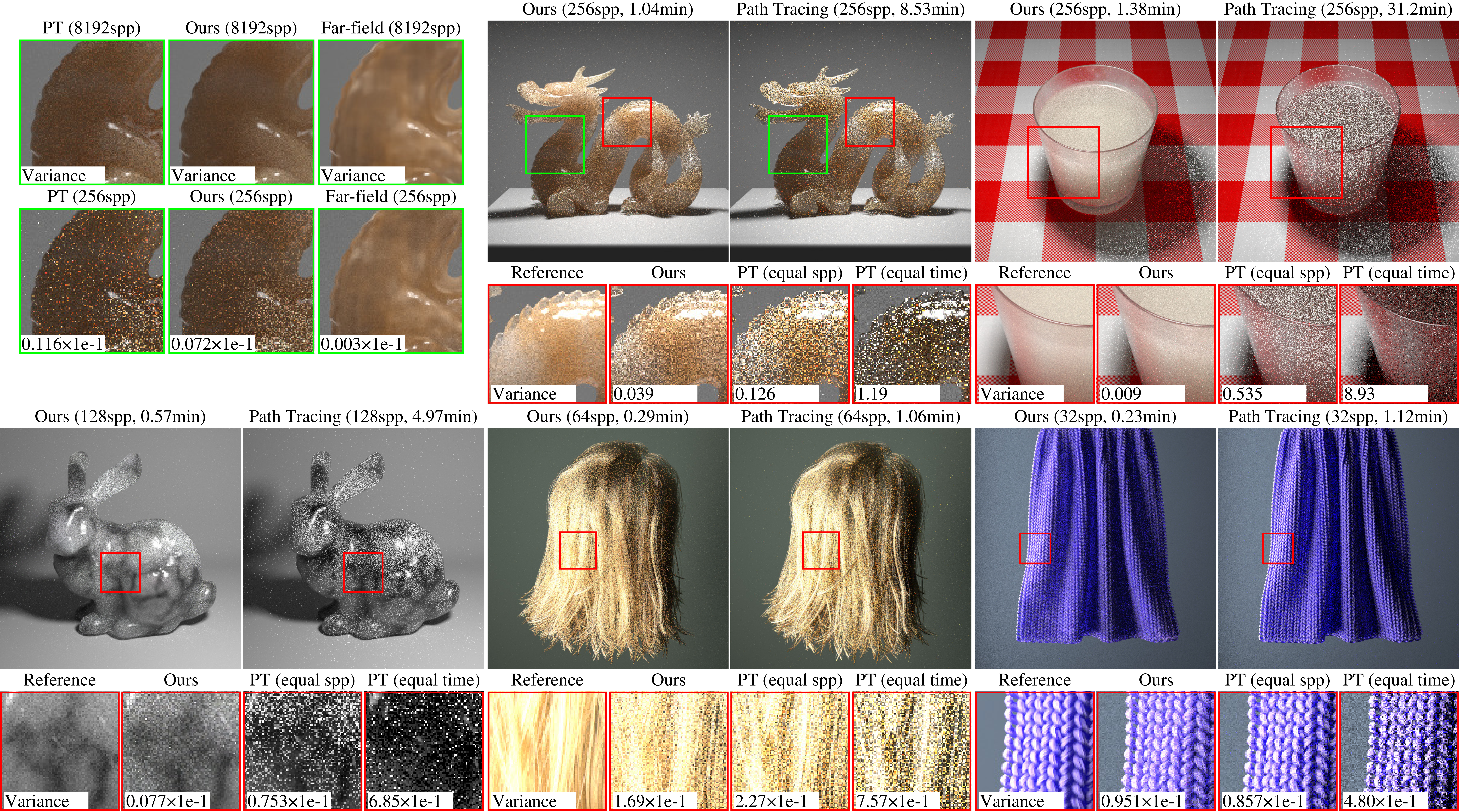}
    \caption{\textbf{Qualitative rendering variance comparison.}
    \methodA{} demonstrates the least variance but its rendering does not match the path-traced reference (top left).
    Overall, our method achieves less variance than path tracing in both equal spp and equal time rendering with comparable rendering quality. The numbers show the variance of the insets.
    }
    \label{fig:4-efficiency}
\end{figure*}
\begin{figure*}[t]
    \centering
     \setlength\tabcolsep{2.5 pt}
    \resizebox{0.99\linewidth}{!}{
    \begin{tabular}{cc}
     spp vs. log variance & time (minutes) vs. log variance\\[1.5pt]
    \includegraphics[width=0.55\linewidth]{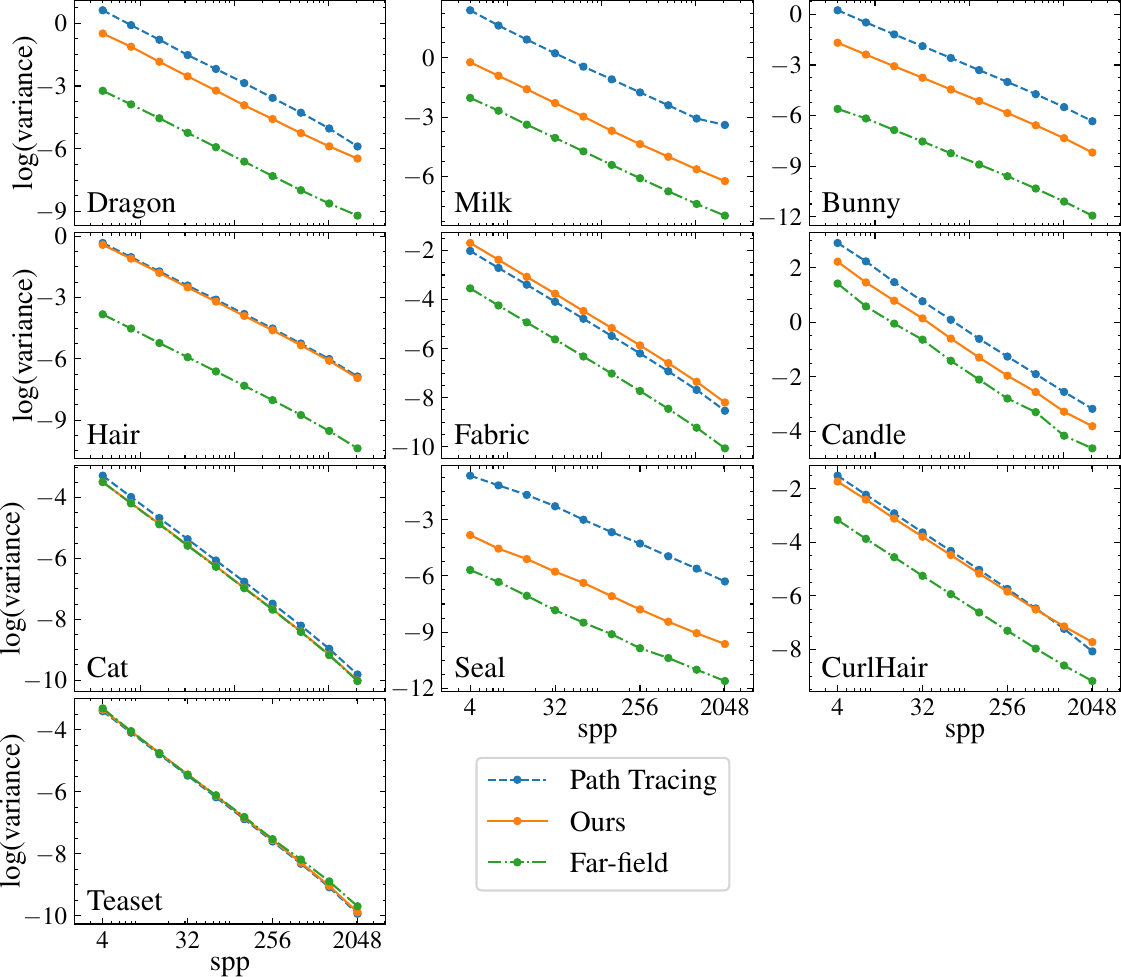}&
    \includegraphics[width=0.55\linewidth]{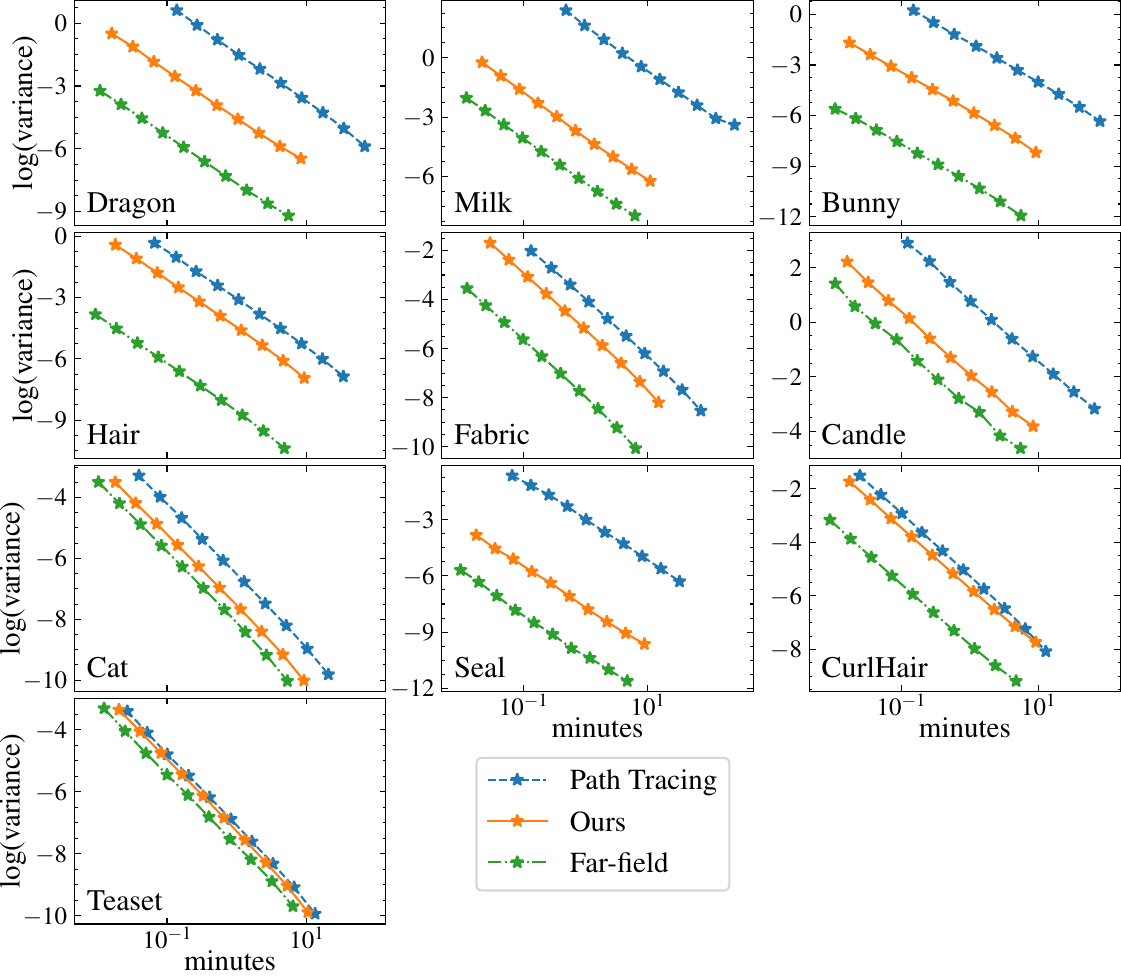}
    \end{tabular}
    }
    \caption{\textbf{Convergence graphs.} Columns 1-3 show the log-log plot of the rendering variance with respect to the spp. Columns 4-6 show the log-log plot of the variance with respect to the rendering time. Our method achieves better variance reduction than path tracing in both comparisons, and the improvement is more significant on more complex scenes.
    The variance is computed with respect to a high spp rendering for each method separately \textit{without} considering reconstruction accuracy (\cref{tab:4-accuracy}).
    \methodA{} does converge faster but is not consistent with the path-traced ground truth (see \cref{tab:4-accuracy} and \cref{fig:4-acc}).
    }
    \label{fig:4-plot}
\end{figure*}
\subsection{Results}
\label{subsec:4-results}
\paragraph{Accuracy of light transport.}
Figure~\ref{fig:4-acc} shows each asset rendered under area or environment light with background geometry contributing indirect illumination.
A rendering of multiple assets composed into a single scene is provided in the supplement.
Although this is a near-field setup,
the diffuse background essentially smooths the illumination,
so the far-field baseline still provides a reasonable approximation of the path-traced reference.
On closer inspection, however, the 6D light transport that ignores $\xxi$ overestimates incoming radiance especially near the shadowed region of \textit{Candle}, \textit{Seal}, and \textit{CurlHair};
additional comparisons under purely far-field and purely near-field lighting in the supplement highlight when these artifacts are least and most visible.
Due to the variance issue of the regression loss,
the far-field model also fails to reproduce strong view-dependent effects,
including the interreflections of the \textit{Teaset} and the \textit{Cat} in the mirror, where the cat's interior white-red-blue volumetric slabs appear purple in the front view.
Benefiting from the correct 8D light transport modeling and the low-variance distribution-learning objective,
our model accurately handles these effects and achieves consistently lower MSE with respect to the reference renderings (\cref{tab:4-accuracy}).

% candle, cat, seal hair, teaset
\paragraph{Sampling efficiency.}
Table~\ref{tab:4-efficiency} reports the rendering variance and speed under 128 spp with the influence of rendering bias removed.
%Because the far-field model pre-integrates over $\xxi$,
%it exhibits much less variance than our method and does not need to evaluate an $\xxi$-dependent network, making it around twice faster in inference.
Because the far-field model pre-integrates over $\xxi$ and does not need to evaluate an $\xxi$-dependent network,
it exhibits much less variance than our method and is around twice faster in inference.
However, this is at the cost of inaccurate light transport as discussed above.
%Compared to path tracing,
%our pre-baking noticeably reduces the variance on assets with volumetric interiors (\cref{fig:4-efficiency}),
%for which naive path tracing cannot exploit emitter sampling to reduce variance.,
Our pre-baking noticeably reduces the variance on assets with volumetric interiors,
for which naive path tracing cannot exploit emitter sampling to reduce variance (\cref{fig:4-efficiency}).
On surface-only assets where the MIS is effective,
we achieve similar variance or are slightly worse due to the lack of MIS over $\xxi$ (\cref{subsec:4-limitation}).
But path-tracing over long scattering paths can be expensive especially for high-albedo assets (\textit{Seal}, \textit{Milk}, and \textit{Hair}),
so we still achieve less variance in equal time rendering.
As shown in the convergence plot (\cref{fig:4-plot}),
the advantage of our method grows with the asset's complexity;
we do not gain too much performance on simple assets like \textit{Teaset} that are already efficient to render with path tracing.

% milk, bunny, dragon, fur, fabric
\begin{table}[t]
    \centering
    \setlength\tabcolsep{8.5 pt}
    \caption{\textbf{Timing of each training stage.}
    Our networks are slower to optimize than the far-field model (including both light transport and the importance sampling module), while our data generation is substantially faster. The time is averaged over all volumetric and surface-only assets separately.
    }
    \resizebox{0.99\linewidth}{!}{
    \begin{tabular}{clccc}
    \toprule
    \multirow{2}{*}{\textbf{Asset type}}&\multirow{2}{*}{\textbf{Method}}& \multicolumn{3}{c}{\textbf{Time (hours$\downarrow$)}}\\
    & & Data generation & Optimization & Total\\
    \midrule
    \multirow{2}{*}{Volumetric}
    &\textbf{\methodA{}}&5.91 &0.84&6.75\\
    &\textbf{Ours}&0.55&1.68&2.23\\
    \midrule
    \multirow{2}{*}{Surface}
    &\textbf{\methodA{}}&1.26&0.72&1.98\\
    &\textbf{Ours}&0.33&1.45&1.78\\
    \bottomrule
    \end{tabular}
    }
    \label{tab:4-training}
\end{table}
\paragraph{Training time.}
As shown in \cref{tab:4-training}, the far-field model is faster to optimize as it mainly evaluates MLPs.
In contrast, pdf evaluations dominate our normalizing flow training,
which does not have an efficient built-in implementation.
However, generating each far-field training data sample requires thousands of spp,
while our data is produced online with only 1 spp.
Taking both factors into account, our full training is $3\times$ faster on volumetric assets and roughly comparable ($1.1\times$) on surface-only assets.
Note our data are resampled for 30 times (\cref{subsec:3-architecture}),
which would make the far-field training take days if using the same strategy.

\subsection{Ablation study}
\label{subsec:4-ablation}

\paragraph{Network architecture.}
Table~\ref{tab:4-architecture} shows the performance trade-off between different variants of our model.
Reducing the network size lowers the rendering time but increases reconstruction error.
The direct–scattering separation improves rendering quality without introducing a noticeable decrease in rendering speed.
All the variants have similar rendering variance.

\begin{table}[t]
    \centering
    \setlength\tabcolsep{5 pt}
    \caption{\textbf{Ablation on network architectures.}
    'Smaller network' decreases the MLP width to 64 in both $\mathbf{p}_{\bm{\theta}}$ and $\bm{\alpha}_{\bm{\theta}}$ and uses 16-knot rqs in normalizing flows, which speeds up the rendering while introducing more reconstruction error. The inference speed is similar without the direct-scattering separation, yet the rendering quality is decreased.
    }
    \resizebox{0.99\linewidth}{!}{
    \begin{tabular}{l c c c}
        \toprule
        \textbf{Model variants} &  \textbf{$100\times$MSE$\downarrow$} & \textbf{$100\times$Variance$\downarrow$} & \textbf{seconds$\downarrow$}\\
        \midrule
        Ours & 0.200 & 0.240 & 12.21\\
        Smaller network & 0.416 & 0.210 & 7.16\\
        Without separation & 0.365 & 0.227 & 12.17\\
        %\midrule
        %Bounding box& 0.221 & 1.405 & 12.14\\
        %Bounding convex hull& 0.231 & 1.191 & 11.49\\
        %Bounding sphere& 0.203 & 1.709 & 11.92\\
        \bottomrule
    \end{tabular}
    }
    \label{tab:4-architecture}
\end{table}
%\begin{figure}[t]
%    \centering
%    \setlength\tabcolsep{0.25 pt}
%    \resizebox{0.99\linewidth}{!}{
%    \begin{tabular}{cccc}
%        Reference & Bounding box & Bounding sphere & Convex hull\\
%         &
%         \includegraphics[width=0.28\linewidth]{images/4-bounding-box.pdf}&
%         \includegraphics[width=0.28\linewidth]{images/4-bounding-sphere.pdf}&
%         \includegraphics[width=0.28\linewidth]{images/4-bounding-convex-hull.pdf}\\
%    \end{tabular}
%    }
%    \caption{Caption}
%    \label{fig:4-bounding-geometry}
%\end{figure}
\begin{figure}[t]
    \centering
    \includegraphics[width=0.99\linewidth]{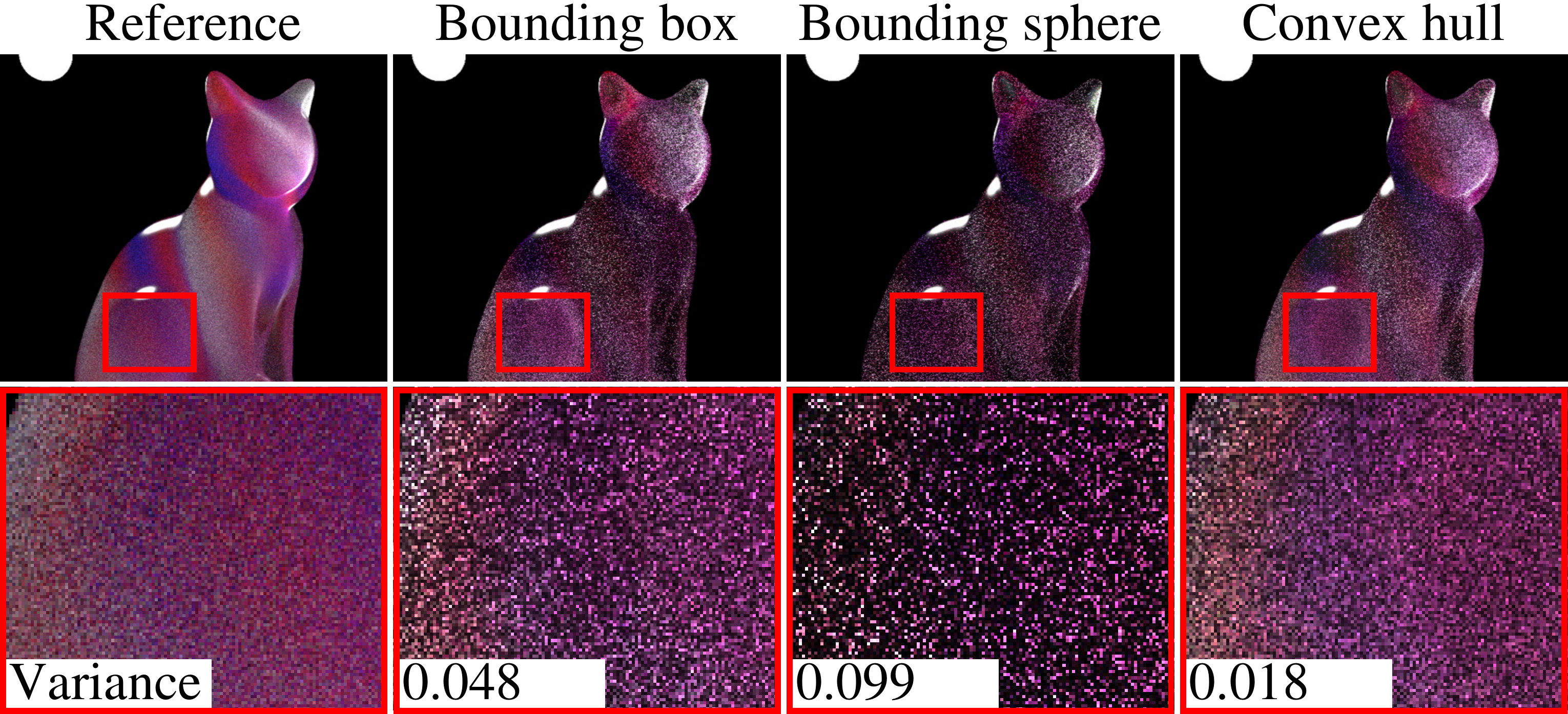}
    \caption{\textbf{Bounding geometry for incident ray parameterization.} Bounding geometries with smaller surface area like convex hull tend to exhibit less variance. The renderings use 128 spp. Intersections on each geometry proxy above are mapped to cylindrical coordinates to fit into the normalizing flow.
    }
    \label{fig:4-bounding-geometry}
\end{figure}
\paragraph{Bounding geometry proxy.}
Besides the bounding box, we can also take intersections on other geometry proxies as the parameterization of incident rays.
As shown in \cref{fig:4-bounding-geometry},
the rendering variance is reduced with tighter bounding geometry like convex hull but increased with bounding sphere that has larger surface area for \textit{Cat} than bounding box.
This is because the variance of $\xxi$-sampling correlates to the surface area of bounding geometry.
Neural assets using bounding boxes already demonstrate low rendering variance in most of our scenes,
so we choose it for the simplicity of its implementation.

\begin{figure}[t]
    \centering
    \includegraphics[width=0.99\linewidth]{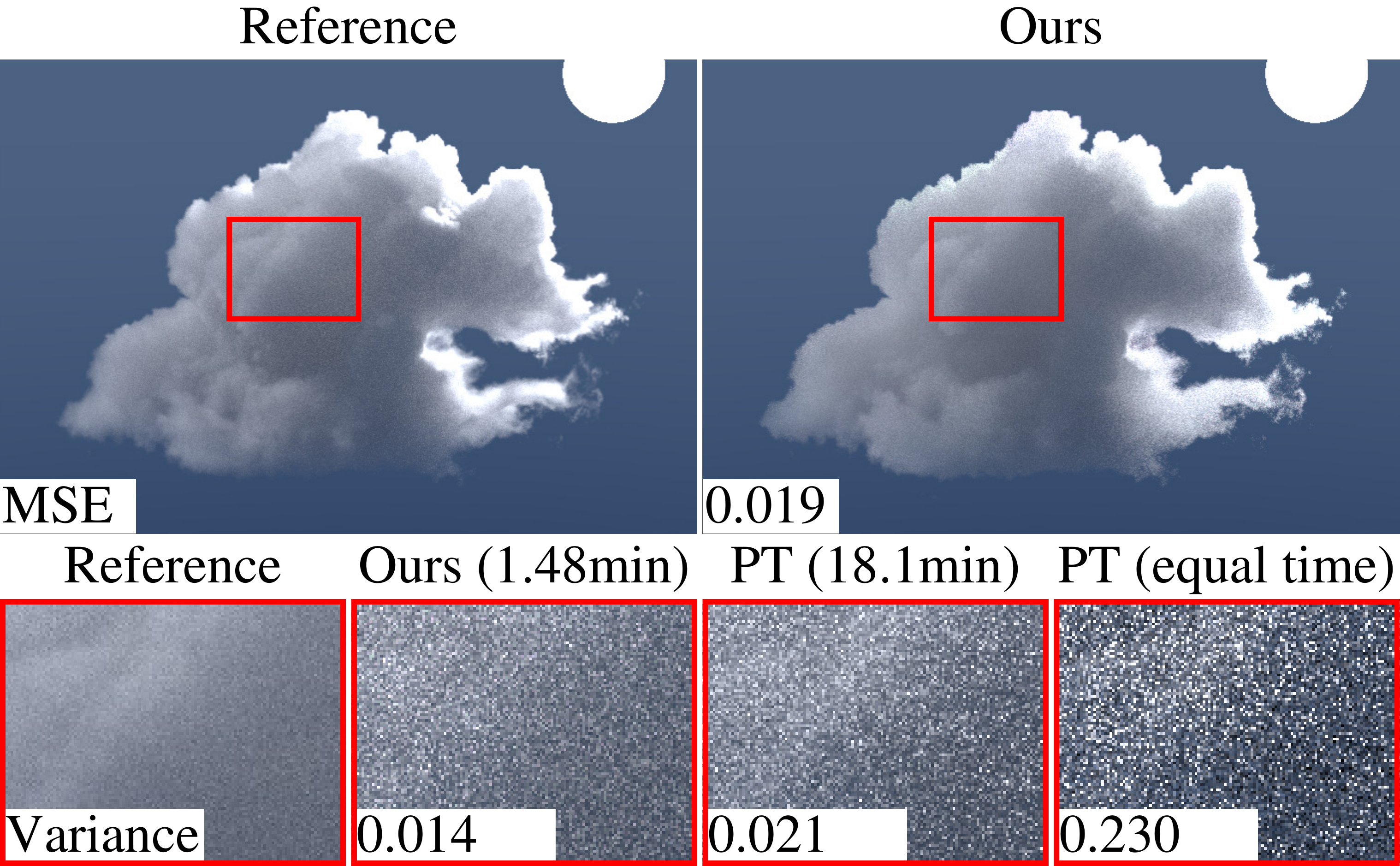}
    \caption{\textbf{Our method extended to a purely volumetric asset.}
    The \textit{Cloud} is illuminated by a sphere area light and an environment light.
    Our rendering closely matches the path-traced reference while it achieves similar equal-spp variance but faster inference (insets rendered at 128 spp).
    }
    \label{fig:4-volume}
\end{figure}
\paragraph{Extension to purely volumetric asset.}
For assets without a solid boundary,
we can define $\xxo$ by transmittance-sampling the location of the first scattering event.
Our model then learns the remaining light transport inside the volume.
Figure~\ref{fig:4-volume} shows the \textit{Cloud} can be reconstructed using this approach.
We do not observe significant variance reduction with our pre-baking,
as path tracing can apply MIS to the pure volume.
However, our model is faster to evaluate than explicit multi-scattering,
which leads to improved equal-time variance compared to path tracing.

\begin{figure}[t]
    \centering
    \setlength\tabcolsep{0.5 pt}
    \resizebox{0.99\linewidth}{!}{
    \begin{tabular}{cc}
        Outside convex hull $\checkmark$ & Inside convex hull $\times$\\
        \includegraphics[width=0.55\linewidth]{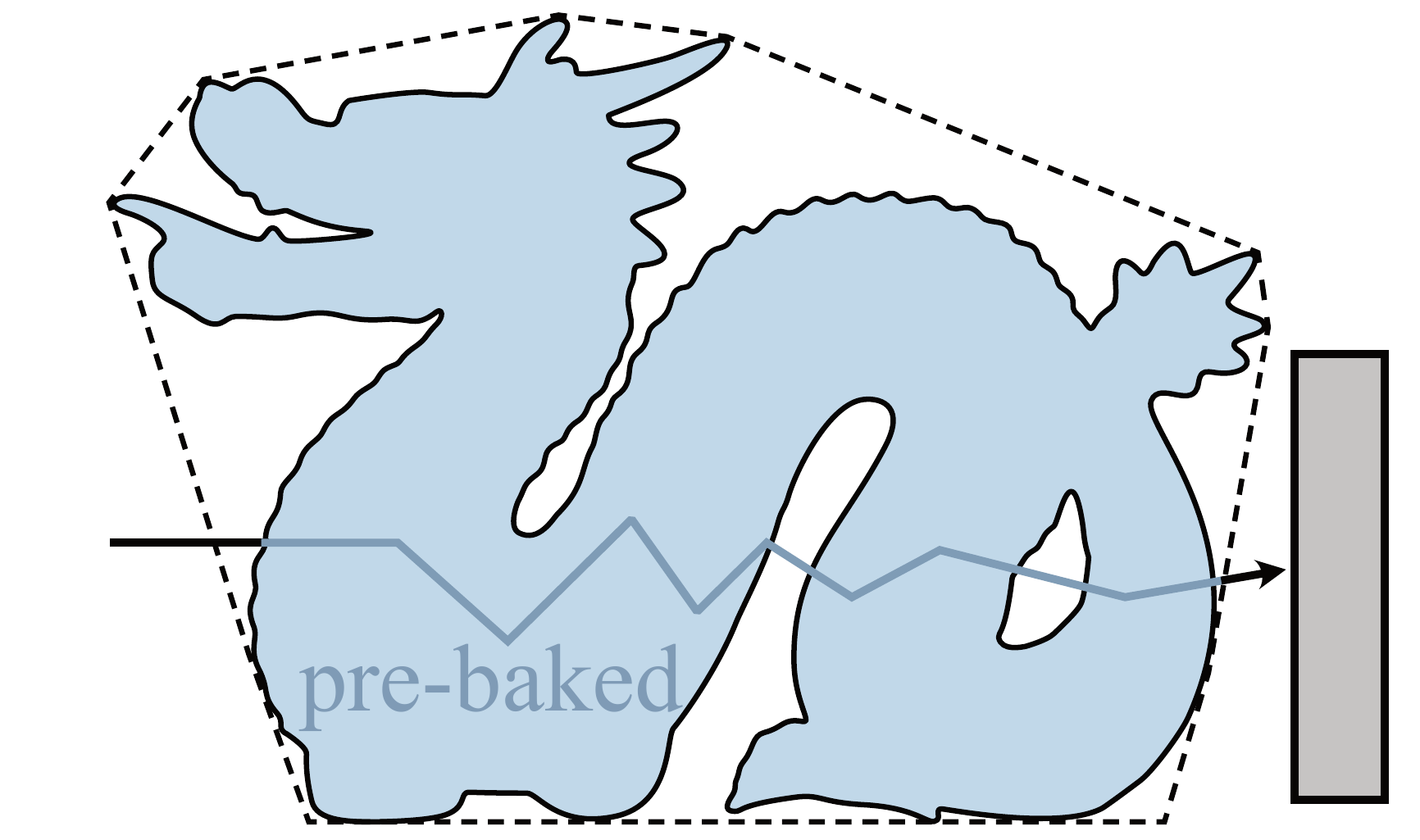}&
        \includegraphics[width=0.55\linewidth]{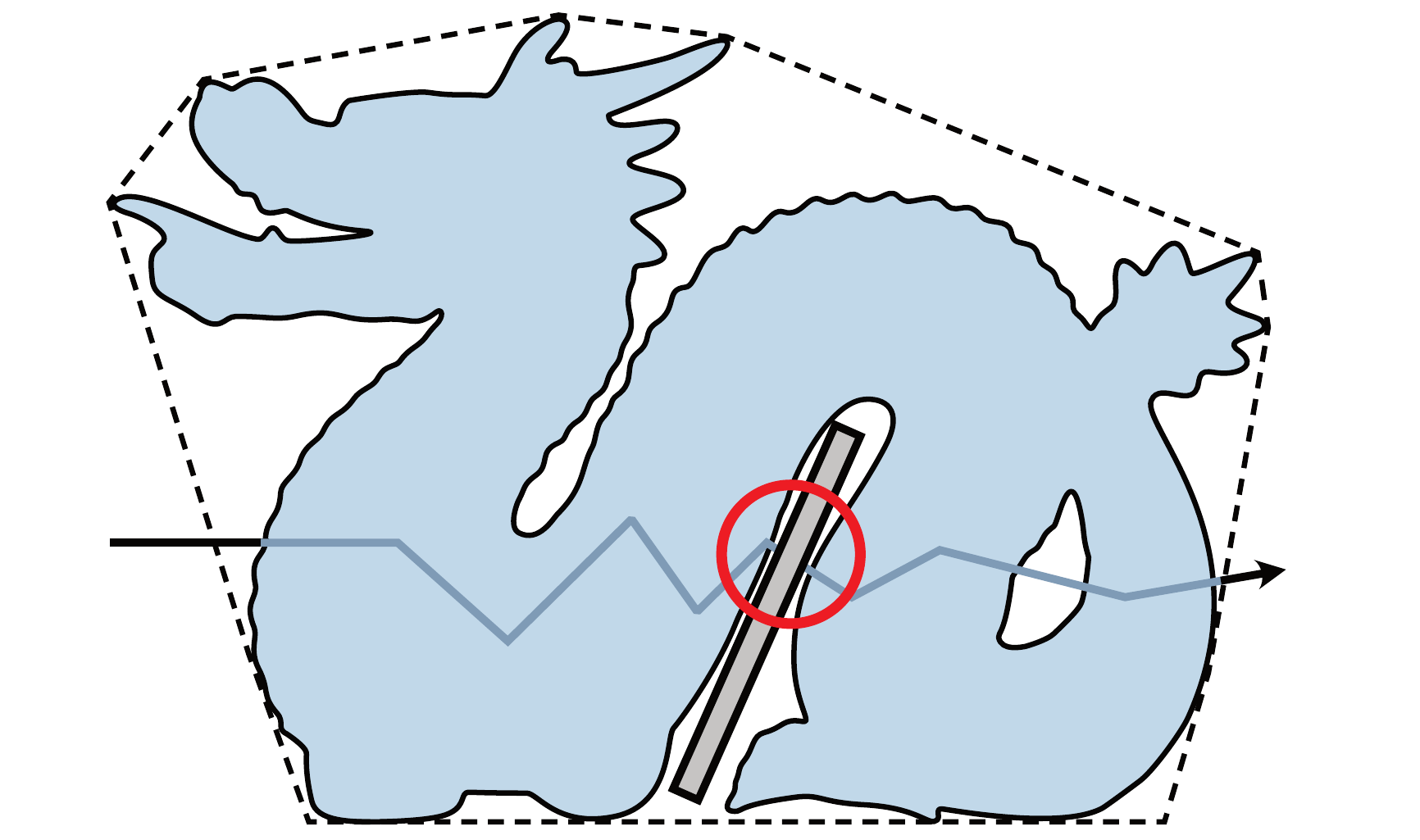}\\
        Path Tracing & Ours\\
        \includegraphics[width=0.55\linewidth]{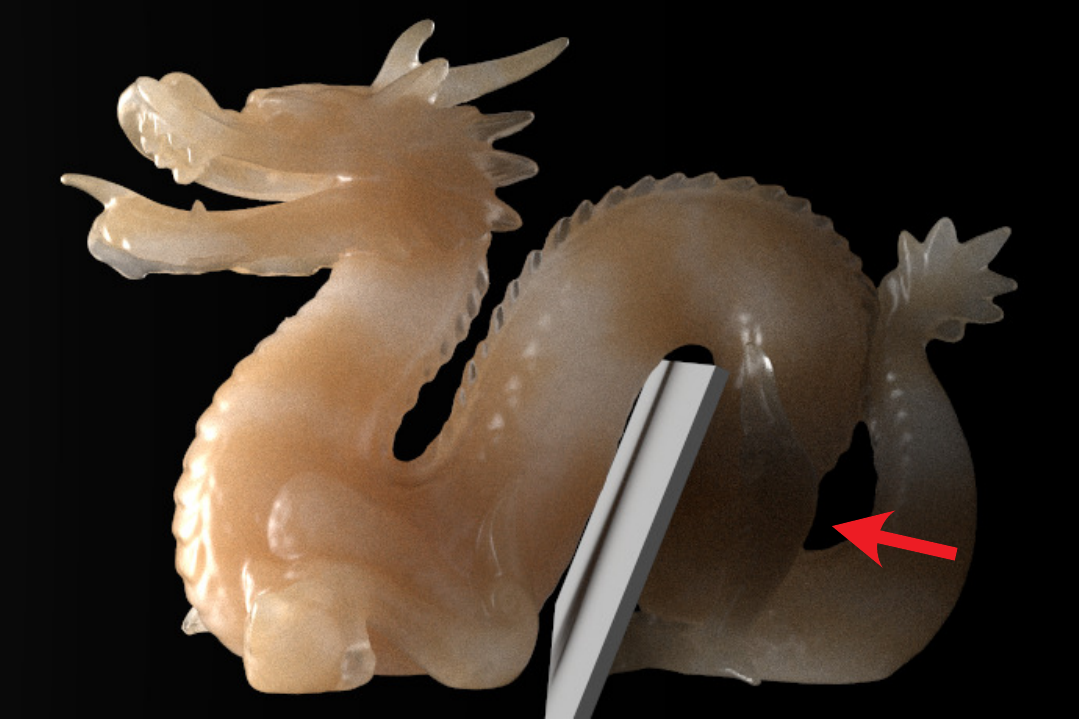}&
        \includegraphics[width=0.55\linewidth]{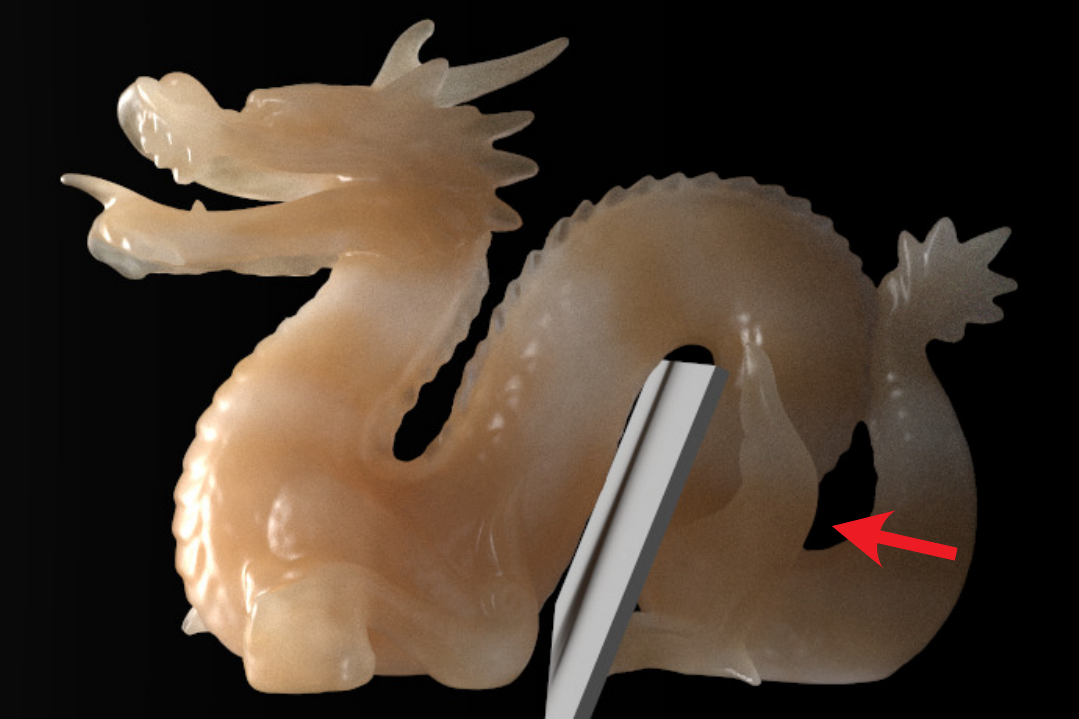}\\
    \end{tabular}
    }
    \caption{
    \textbf{Failure case of our model.}
    An area emitter is placed on the right of \textit{Dragon}.
    The diffuse slab blocks the pre-baked light paths (top right), causing our model to miss occlusion effects (red arrow).
    Our method assumes no additional geometry lies inside the asset's convex hull (top left).
    }
    \label{fig:4-limitation}
\end{figure}
\subsection{Limitations}
\label{subsec:4-limitation}
We pre-bake light transport for each asset in isolation,
so the internal scattering is essentially bounded by the asset's convex hull.
When inserting the asset into a new scene,
our model remains accurate only if the internal scattering structure does not change.
This means no other objects can intersect with the asset's convex hull (\cref{fig:4-limitation}).
Applying convex decomposition to the asset then pre-baking each component separately can resolve this issue.
Meanwhile, we currently do not use MIS over $\xxi$,
which may lead to high rendering variance under small emitters like point lights.
Sampling incident locations visible to the emitters may help.

The usage of normalizing flows also introduces limitations.
On the one hand, the normalizing flow evaluation is less efficient than MLP,
so an MLP-based light transport model remains preferable if the far-field approximation is sufficiently accurate.
On the other hand, 
normalizing flows have an inductive bias toward smooth distributions,
making it difficult to reproduce high-frequency phenomena,
such as specular interreflections between complex shapes or caustics and glints that lie in a sub-manifold of the incident domain.
\section{Conclusion and Future Work}
\label{sec:conclusion}
We have demonstrated an 8D neural representation that learns global light transport inside 3D assets from forward path-traced samples.
This distribution-learning framework provides low-variance gradient estimates for optimizing our light transport model,
while the 8D parameterization accurately pre-bakes internal scattering even under near-field illumination.
Together, our method noticeably reduces rendering variance and time compared to standard path tracing while preserving high visual fidelity.
Looking forward, the model has the potential to deliver fast and accurate global illumination in real-time applications.

\begin{acks}
This work was supported in part by NSF grants 2212085, 2341952, 2105806,
NSF Chase-CI grants 2100237, 2120019,
and ONR grant N00014-23-1-2526.
We also acknowledge an NVIDIA Fellowship,
gifts from Adobe, Google and Qualcomm, the Ronald L. Graham
Chair and the UC San Diego Center for Visual Computing.
Ramamoorthi acknowledges a part-time appointment at NVIDIA.
\end{acks}

\bibliographystyle{ACM-Reference-Format-few-warning}
\bibliography{bibliography}

\begin{appendix}
    \section{Additional Derivations}

\subsection{Isotropic assumption of NeuralSSS}
NeuralSSS~\cite{tg2024neural} utilizes an MLP $\mathbf{S}_{\bm{\theta}}$
to predict the subsurface scattering at $(\xxo,\omegao)$ given the input of $N$ incident rays $(\bx_n,\bo_n)$ with radiance $\capl_n=\capl_i(\bx_n,\bo_n)$.
To train the network,
\citeauthor{tg2024neural} first randomly sample $(\xxo,\omegao)$ queries then path-sample $N$ sub-rays for each outgoing ray,
recording their configuration $(\bx_n,\bo_n)$ and throughput $\bm{\beta}_n$ when leaving the subsurface media;
note this is different from our sampling that records the incident configuration when leaving the entire asset.
Each sub-ray is further paired with a random incident radiance value $\capl_n$ to compute the loss:
\begin{equation}
\left(\mathbf{S}_{\bm{\theta}}(\xxo,\omegao,
\{\bx_n,\bo_n,\capl_n\}_{n=1}^N)
-\frac{1}{N}
\sum_{n=1}^N
\bm{\beta}_n\capl_n
\right)^2.
\label{eq:6-neuralsss}
\end{equation}%
Unlike the far-field regression loss in Eq.~(3) of the paper, 
\cref{eq:6-neuralsss} here uses the same $(\bx_n,\bo_n)$ for network queries and ground truth estimation.
This correlation deviates $\mathbf{S}_{\bm{\theta}}$ from learning the true outgoing radiance.
Let $q(\bx_n,\bo_n|\xxo,\omegao)=q_n$ denote the marginal distribution of path-sampled incident ray (note the path-sampling distribution may differ from the scattering distribution $\mathbf{p}$) and assume $N=1$,
the expectation of the loss gradient above is:
\begin{gather}
2\iint
q_n
\left(\mathbf{S}_{\bm{\theta}}-\bm{\beta}_n\capl_n\right)
\nabla_{\bm{\theta}}\mathbf{S}_{\bm{\theta}}
\mathrm{d}\bx_n\mathbf{d}\bo_n
\notag\\
=2\iint q_n \mathbf{S}_{\bm{\theta}}\nabla_{\bm{\theta}}\mathbf{S}_{\bm{\theta}}\mathrm{d}\bx_n\mathbf{d}\bo_n
-2\iint q_n\bm{\beta}_n\capl_n\nabla_{\bm{\theta}}\mathbf{S}_{\bm{\theta}}\mathrm{d}\bx_n\mathbf{d}\bo_n\notag.
\end{gather}
The second integral equals $\capl_{o}(\xxo,\omegao;\capl_i\nabla_{\bm{\theta}}\mathbf{S}_{\bm{\theta}})$ as discussed in Sec.~3.1 of the paper;
writing it in the global light transport form (Eq.~(1) of the paper) gives:
\begin{equation}
\begin{gathered}
2\iint q_n \mathbf{S}_{\bm{\theta}}\nabla_{\bm{\theta}}\mathbf{S}_{\bm{\theta}}\mathrm{d}\bx_n\mathbf{d}\bo_n
-2\iint \capf_n\capl_n\nabla_{\bm{\theta}}\mathbf{S}_{\bm{\theta}}\mathrm{d}\bx_n\mathbf{d}\bo_n
\\
=2\iint q_n
\left(\mathbf{S}_{\bm{\theta}}-\frac{\capf_n\capl_n}{q_n}\right)\nabla_{\bm{\theta}}\mathbf{S}_{\bm{\theta}}
\mathrm{d}\bx_n\mathbf{d}\bo_n
\\
=\nabla_{\bm{\theta}}\iint
q_n\left(
\mathbf{S}_{\bm{\theta}}-\frac{\capf_n\capl_n}{q_n}
\right)^2\mathrm{d}\bx_n\mathbf{d}\bo_n,
\;\capf_n=\capf(\xxo,\omegao,\xxi,\omegai).
\end{gathered}
\end{equation}%
This suggest $\mathbf{S}_{\bm{\theta}}$ regresses towards the estimator $\frac{\capf_n\capl_n}{q_n}$ rather than the estimator expectation (outgoing radiance) at convergence.
A similar derivation can be made for arbitrary $N$,
where the network learns $\frac{1}{N}\sum_{n=1}^{N}\bm{\beta}_n\capl_n$.
Therefore, it is the expectation of $\mathbf{S}_{\bm{\theta}}$ under joint probability $q_1\cdots q_n$ that estimates the outgoing radiance:
\begin{equation}
\begin{split}
\int\!\!\cdots\!\!\int
&\prod_{n=1}^N
q_n
\mathbf{S}_{\bm{\theta}}(\xxo,\omegao,
\{\bx_n,\bo_n,\capl_n\}_{n=1}^N)
\mathrm{d}\bx_1\mathrm{d}\bo_1
\cdots\mathrm{d}\bx_n\mathrm{d}\bo_n\\
&=\int\!\!\cdots\!\!\int
\prod_{n=1}^N
q_n\frac{1}{N}\sum_{n=1}^N\frac{\capf_n\capl_n}{q_n}
\mathrm{d}\bx_1\mathrm{d}\bo_1
\cdots\mathrm{d}\bx_n\mathrm{d}\bo_n\\
\\
&=\frac{1}{N}\sum_{n=1}^N
\iint \capf(\xxo,\omegao,\bx_n,\bo_n)\capl_i(\bx_n,\bo_n)\mathrm{d}\bx_n\mathrm{d}\bo_n\\
&=\iint\capf(\xxo,\omegao,\xxi,\omegai)\capl_i(\xxi,\omegai)\mathrm{d}\xxi\mathrm{d}\omegai
\end{split}
\label{eq:6-neuralsss-integral}
\end{equation}%

In order to estimate the integral of $\mathbf{S}_\mathbf{\theta}$ in \cref{eq:6-neuralsss-integral},
a sampling strategy is needed to approximate the path sampling distribution $q$.
\citeauthor{tg2024neural} learn to sample the marginal $q(\xxi|\xxo,\omegao)=\int q(\xxi,\omegai|\xxo,\omegao)\mathrm{d}\omegai$ 
using a normalizing flow $q_{\bm{\theta}}(\xxi|\xxo,\omegao)$
but sample $\omegai$ only by cosine-hemisphere sampling.
Such integral estimator is accurate only if the conditional distribution $q(\omegai|\xxi,\xxo,\omegao)$ is cosine-hemisphere,
implying the subsurface scattering is nearly isotropic under an infinite slab.
Note $q(\omegai|\xxi,\xxo,\omegao)$ also depends on the boundary shape of subsurface scattering,
so even if we use cosine-hemisphere path sampling for training,
the conditional distribution will still not be cosine-hemisphere as long as the boundary is not an infinite slab.
Constructing a cosine-hemisphere $q(\omegai|\xxi,\xxo,\omegao)$ over arbitrary boundary shape is generally difficult.

\begin{figure}[t]
    \centering
    \includegraphics[width=0.99\linewidth]{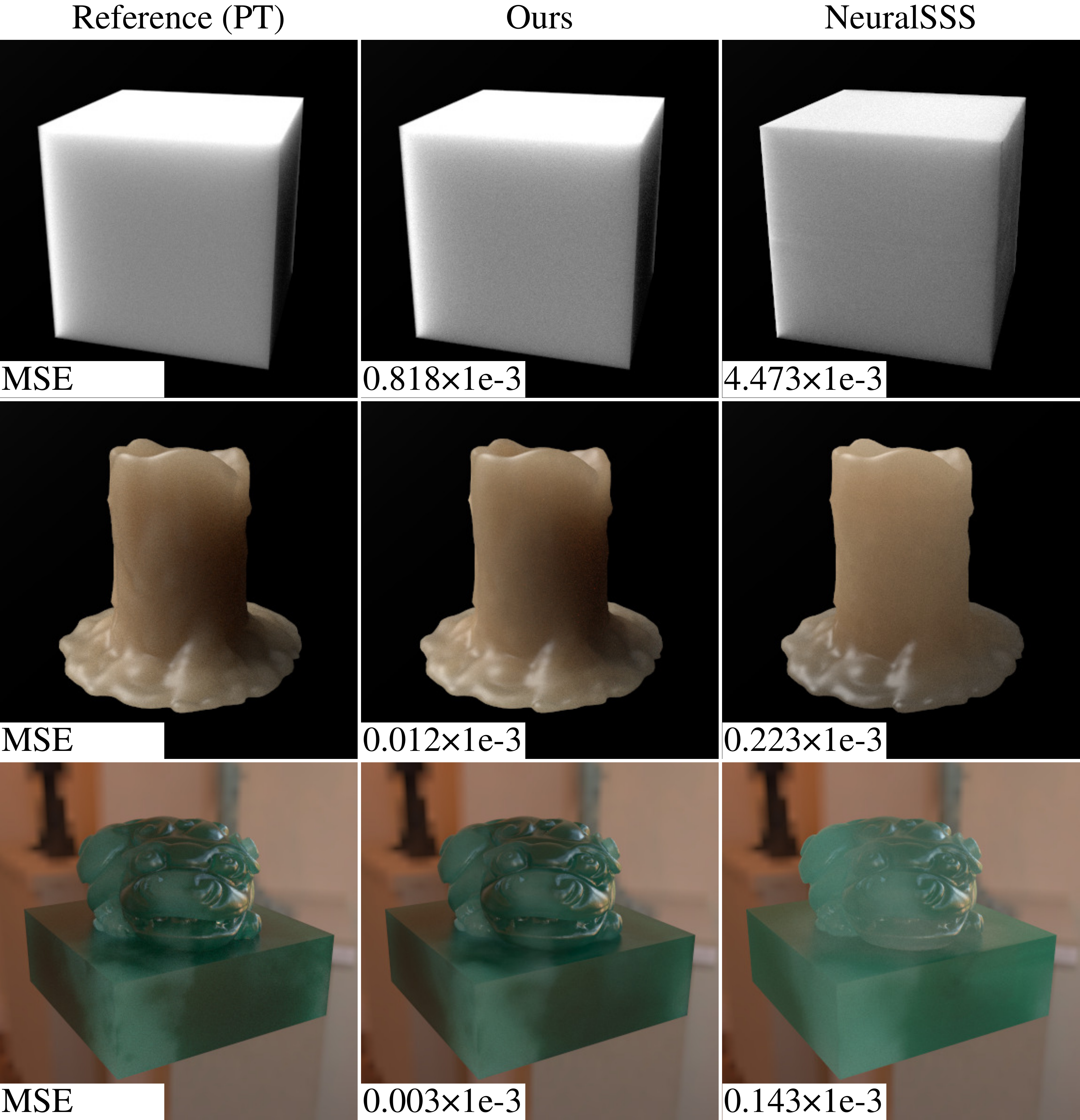}
    \caption{\textbf{Comparison with NeuralSSS.}
    Images are rendered using an environment light at 4096 spp.
    NeuralSSS gives a reasonable approximation of \textit{Cube}, but errors can be seen near edges where the scattering is no longer isotropic.
    For more complex assets presented in the paper (\textit{Candle} and \textit{Seal}), NeuralSSS fails even under far-field lighting.
    }
    \label{fig:6-neuralsss}
\end{figure}
\paragraph{Comparison with NeuralSSS.}
It can be seen in \cref{fig:6-neuralsss} that NeuralSSS gives reasonable approximation of the subsurface scattering inside a homogeneous \textit{Cube} under isotropic internal scattering without dielectric boundary.
However, it fails to reconstruct the assets shown in our experiment that contains view-dependent scatterings.
Our NeuralSSS model uses $N=32$ as suggested by \citeauthor{tg2024neural}.
%To prevent an exponential increase of sub-rays:
%each sub-ray will spawn another $N$ rays when re-hitting the subsurfuace medias,
%we only further trace one sub-ray and set $\capl_n=0$ for the rest then multiplying the result by $N$.

\subsection{Projection factors of the reparameterizations}
We use the fact that $p(\mathbf{x})\!=\!p(\mathbf{y})\frac{|\mathrm{d}\mathbf{y}|}{|\mathrm{d}\mathbf{x}|}$ for a distribution $p(\mathbf{x})$ and its reparameterization $\mathbf{y}(\mathbf{x})$.
The area of $\mathrm{d}\mathbf{u}_i$ and $\mathrm{d}\xxi$ projected along direction $\omegai$ are equal, so we have:
\begin{equation}
\begin{split}
    |\mathrm{d}\mathbf{u}_i||\mathbf{n}_{\mathbf{u}_i}\cdot\omegai|=
    |\mathrm{d}\xxi||\mathbf{n}_{\mathbf{u}_i}\cdot\omegai|
    \Rightarrow
    \frac{|\mathrm{d}\mathbf{u}_i|}{|\mathrm{d}\xxi|}=\frac{|\mathbf{n}_{\xxi}\cdot\omegai|}{|\mathbf{n}_{\mathbf{u}_i}\cdot\omegai|}.
\end{split}
\end{equation}
$\mathbf{u}_i$ on the zero-centered bounding box can be converted to solid angle domain by $\bm{\omega}=\frac{\mathbf{u}_i}{\sqrt{\mathbf{u}_i\cdot\mathbf{u}_i}}$ with $\frac{|\mathrm{d}\bm{\omega}|}{|\mathrm{d}\mathbf{u}_i|}=\frac{|\mathbf{n}_{\mathbf{u}_i}\cdot\bm{\omega}|}{\mathbf{u}_i\cdot\mathbf{u}_i}$,
and since projection factor between a solid angle and cylindrical coordinate is always 1, we have:
\begin{equation}
    \frac{|\mathrm{d}\mathbf{s}|}{|\mathrm{d}\mathbf{u}_i\mathrm{d}\omegai|}
    =\frac{|\mathrm{d}\bm{\omega}|}{|\mathrm{d}\mathbf{u}_i|}=\frac{|\mathbf{n}_{\mathbf{u}_i}\cdot\bm{\omega}|}{\mathbf{u}_i\cdot\mathbf{u}_i}=\frac{|\mathbf{n}_{\mathbf{u}_i}\cdot\mathbf{u}_i|}{\left(\sqrt{\mathbf{u}_i\cdot\mathbf{u}_i}\right)^3}.
\end{equation}

\subsection{Path tracing details}
\label{subsec:6-pathtracing}
We describe the algorithm in the context of a surface-based path tracer, though the same idea extends naturally to volumetric path tracing.
For each camera ray, we perform standard path tracing for multiple bounces,
accumulating pixel radiance $\capl$ and throughput $\bm{\beta}$.
When the ray hits a neural asset,
the total contribution to the pixel can be expressed as $\capl+\bm{\beta}\capl_o$, where $\capl_o$ is given by Eq.~1 of the paper.
We reparameterize $\xxi$ to $\mathbf{u}_i$,
so that the resulting $\mathbf{u}_i$-integral can be estimated by sampling $\mathbf{u}_i\sim \mathbf{p}^c_{\bm{\theta}}(\mathbf{u}_i|\xxo,\omegao)=\mathbf{p}^c_{\mathbf{u}_i}$ over a random color channel $c$ ($\xxo,\omegao$ are omitted for simplicity):
\begin{equation}
\begin{split}
    \capl_o &= \iint\!\!\capf_{\bm{\theta}}(\mathbf{u}_i,\omegai)\capl_i(\mathbf{x}_i,\omegai)\mathrm{d}\mathbf{u}_i\mathrm{d}\omegai\\
    &=\mathop{\mathbb{E}}_{\mathbf{u}_i\sim\mathbf{p}^c_{\mathbf{u}_i}}
\left[
\int\frac{\capf_{\bm{\theta}}(\mathbf{u}_i,\omegai)}{
\mathbf{p}^c_{\mathbf{u}_i}(\mathbf{u}_i)
}\capl_i(\mathbf{x}_i,\omegai)\mathrm{d}\omegai
\right]\\
&\quad\quad\quad
\capf_{\bm{\theta}}(\mathbf{u}_i,\omegai)
=\mathbf{F}_{\bm{\theta}}(\xxi,\omegai)\left\vert\tfrac{
\mathbf{n}_{\mathbf{u}_i}\cdot\omegai}{
\mathbf{n}_{\xxi}\cdot\omegai
}\right\vert.
\end{split}
\end{equation}
The incident radiance $\capl_i(\mathbf{x}_i,\omegai)$ is decomposed into surface emission $\capl_e(\mathbf{x}_i',-\omegai)$ and outgoing radiance $\capl_o(\mathbf{x}'_i,-\omegai)$,
where $\xxi'$ is the surface intersection of ray $(\xxi,\omegai)$:
\begin{equation}
\begin{split}
\mathop{\mathbb{E}}_{\mathbf{u}_i\sim\mathbf{p}^c_{\mathbf{u}_i}}
&\left[\int\frac{\capf_{\bm{\theta}}(\mathbf{u}_i,\omegai)}{
\mathbf{p}^c_{\mathbf{x}_i}(\mathbf{u}_i)
}\capl_i(\mathbf{x}_i,\omegai)\mathrm{d}\omegai\right]\\
&=\mathop{\mathbb{E}}_{\mathbf{u}_i\sim\mathbf{p}^c_{\mathbf{u}_i}}
\left[\int\frac{\capf_{\bm{\theta}}(\mathbf{u}_i,\omegai)}{
\mathbf{p}^c_{\mathbf{u}_i}(\mathbf{u}_i)
}\capl_e(\mathbf{x}'_i,-\omegai)\mathrm{d}\omegai\right]\\
&+\mathop{\mathbb{E}}_{\mathbf{u}_i\sim\mathbf{p}^c_{\mathbf{u}_i}}
\left[\int\frac{\capf_{\bm{\theta}}(\mathbf{u}_i,\omegai)}{
\mathbf{p}^c_{\mathbf{u}_i}(\mathbf{u}_i)
}\capl_o(\mathbf{x}'_i,-\omegai)\mathrm{d}\omegai\right].
\end{split}
\end{equation}
The first integral is estimated using MIS between an emitter-sampling distribution (in solid angle domain) $p_e(\omegai)$ and our path-sampling distribution $\mathbf{p}_{\bm{\theta}}^c(\omegai|\mathbf{u}_i,\xxo,\omegao)=\mathbf{p}^c_{\omegai}$:
\begin{equation}
\begin{split}
&\mathop{\mathbb{E}}_{\mathbf{u}_i\sim\mathbf{p}^c_{\mathbf{u}_i}}
\left[\int\frac{\capf_{\bm{\theta}}(\mathbf{u}_i,\omegai)}{
\mathbf{p}^c_{\mathbf{u}_i}(\mathbf{u}_i)
}\capl_e(\mathbf{x}'_i,-\omegai)\mathrm{d}\omegai\right]\\
&=\mathop{\mathbb{E}}_{\mathbf{u}_i\sim\mathbf{p}^c_{\mathbf{u}_i},\bm{\omega}_i\sim p_e}\left[w_{emit}(\omegai)\frac{\capf_{\bm{\theta}}(\mathbf{u}_i,\omegai)}{\mathbf{p}_{\mathbf{u}_i}^c(\mathbf{u}_i)p_e(\omegai)}\capl_e(\xxi',-\omegai)
\right]\\
&+\mathop{\mathbb{E}}_{\mathbf{u}_i\sim\mathbf{p}^c_{\mathbf{u}_i},\bm{\omega}_i\sim \mathbf{p}^c_{\omegai}}\left[w_{path}(\omegai)\frac{\capf_{\bm{\theta}}(\mathbf{u}_i,\omegai)}{\mathbf{p}_{\mathbf{u}_i}^c(\mathbf{u}_i)\mathbf{p}^c_{\omegai}(\omegai)}\capl_e(\mathbf{x}'_i,-\omegai)
\right]\\
&=\capl_{emit} + \capl_{path}.
\end{split}
\end{equation}
We include the emitter visibility to $\capl_e$ and use the power heuristic~\cite{veach1995optimally} with weights $w_{emit}=\frac{p_e^2}{p_e^2+(\mathbf{p}^c_{\omegai})^2}$, $w_{path}=1-w_{emit}$.
The remaining integral is estimated using the same path-sampled direction,
with $\capl_o(\xxi',\omegai)$ further traced in the next bounce:
\begin{equation}
\begin{split}
\mathop{\mathbb{E}}_{\mathbf{u}_i\sim\mathbf{p}^c_{\mathbf{u}_i}}
&\left[\int\frac{\capf_{\bm{\theta}}(\mathbf{u}_i,\omegai)}{
\mathbf{p}^c_{\mathbf{u}_i}(\mathbf{u}_i)
}\capl_o(\mathbf{x}'_i,-\omegai)\mathrm{d}\omegai\right]=\capl_{next}\\
&=\mathop{\mathbb{E}}_{\mathbf{u}_i\sim\mathbf{p}^c_{\mathbf{u}_i},\omegai\sim\mathbf{p}^c_{\omegai}}
\left[\frac{\capf_{\bm{\theta}}(\mathbf{u}_i,\omegai)}{
\mathbf{p}^c_{\mathbf{u}_i}(\mathbf{u}_i)\mathbf{p}^c_{\omegai}(\omegai)
}\capl_o(\mathbf{x}'_i,-\omegai)\right].
\end{split}
\end{equation}
dAs in standard path tracing,
$\capl_{emit}$ is evaluated at the current bounce and accumulated as $\capl\leftarrow\capl+\bm{\beta}\capl_{emit}$.
The evaluation of $\capl_{path}$ is deferred:
we update the throughput $\bm{\beta}\leftarrow\bm{\beta}\times\frac{\capf_{\bm{\theta}}(\mathbf{u}_i,\omegai)}{
\mathbf{p}^c_{\mathbf{u}_i}(\mathbf{u}_i)\mathbf{p}^c_{\omegai}(\omegai)
}$,
project $\mathbf{u}_i$ back to the asset boundary to obtain $\xxi$,
and continue tracing the ray $(\xxi,\omegai)$;
at the next bounce, we update the radiance with the path contribution $\capl\leftarrow \capl + w_{path}(\omegai)\bm{\beta}\capl_e(\xxi',-\omegai)$.

\paragraph{Modifications with direct-scattering separation.}
When $\capf_{\bm{\theta}}$ only models indirect scattering,
the outgoing radiance at $\xxo$ becomes:
\begin{equation}
\begin{split}
    \capl_o&=\!\iint\!\!\capf_{\bm{\theta}}(\mathbf{u}_i,\omegai)\capl_i(\xxi,\omegai)\mathrm{d}\mathbf{u}_i\mathrm{d}\omegai\\
    &+ \!\int\!\!\mathbf{f}(\xxo,\omegao,\omegai')V(\xxo,\omegai')\capl_i(\xxo,\omegai')\mathrm{d}\omegai',
\end{split}
\end{equation}
where the second integral can be estimated in a similar way without needing to sample $\mathbf{u}_i$:
\begin{equation}
\begin{split}
\int&\mathbf{f}(\xxo,\omegao,\omegai')V(\xxo,\omegai')\capl_i(\xxo,\omegai')\mathrm{d}\omegai'\\
&=\mathop{\mathbb{E}}_{\omegai'\sim p_e}\left[
w_{emit}(\omegai')
\frac{\mathbf{f}(\xxo,\omegao,\omegai')}{
p_e(\omegai')
}V(\xxo,\omegai')\capl_e(\xxo',-\omegai')
\right]\\
&+\mathop{\mathbb{E}}_{\omegai'\sim p_{\mathbf{f}}}\left[
w_{path}(\omegai')
\frac{\mathbf{f}(\xxo,\omegao,\omegai')}{
p_{\mathbf{f}}(\omegai')
}V(\xxo,\omegai')\capl_e(\xxo',-\omegai')
\right]\\
&+\mathop{\mathbb{E}}_{\omegai'\sim p_{\mathbf{f}}}\left[
\frac{\mathbf{f}(\xxo,\omegao,\omegai')}{
p_{\mathbf{f}}(\omegai')
}V(\xxo,\omegai')\capl_o(\xxo',-\omegai')
\right]\\
&= \capl'_{emit} + \capl'_{path} + \capl'_{next}.
\end{split}
\end{equation}
$\xxo'$ is the surface intersection of ray $(\xxo',\omegai')$,
and $p_{\mathbf{f}}(\omegai')$ denotes the path-sampling distribution of $\mathbf{f}$.
We use $\omegai'$ to emphasis this direction is sampled independently of the neural asset sample $\omegai$.
$\capl_{emit}'$ can be simply added to the radiance at the current bounce: $\capl\leftarrow\capl+\bm{\beta}\capl'_{emit}$.
However,$\capl_{path}+\capl_{next}$ and $\capl_{path}'+\capl'_{next}$ cannot be  evaluated simultaneously as only one path sample can be carried to the next bounce.
Instead, we draw a direct sample $(\xxo,\omegai')$ and an indirect sample $(\xxo,\omegai')$ then stochastically choose to continue tracing $\capl_{path}'+\capl_{next}'$ with probability $m(\xxo,\omegai',\xxi,\omegai)$ or with $\capl_{path}+\capl_{next}$ otherwise. 
This leads to a modified throughput update rule:
%The modified throughput update and path-sampling estimator are given by:
\begin{equation}
\bm{\beta}\leftarrow\bm{\beta}\times
\begin{cases}
    \frac{1}{m(\xxo,\omegai',\xxi,\omegai)}\frac{\mathbf{f}(\xxo,\omegao,\omegai')V(\xxo,\omegai')}{p_{\mathbf{f}}(\omegai')} & \text{if select } (\xxo,\omegai')\\
    \frac{1}{1-m(\xxo,\omegai',\xxi,\omegai)}
    \frac{\mathbf{F}_{\bm{\theta}}(\mathbf{u}_i,\omegai)}{\mathbf{p}^c_{\mathbf{u}_i}(\mathbf{u}_i)\mathbf{p}^c_{\omegai}(\omegai)} & \text{otherwise}
\end{cases},
\end{equation}
where $\frac{1}{m}$ and $\frac{1}{1-m}$ are the correction factors of the stochastic selection.
%\begin{equation}
%\capl \leftarrow \capl +
%\begin{cases}
%    w_{path}(\omegai') \bm{\beta} \capl_e(\xxo',-\omegai') & \text{if select } (\xxo,\omegai')\\
%    w_{path}(\omegai) \bm{\beta} \capl_e(\xxi',-\omegai) & \text{otherwise}
%\end{cases},
%\label{eq:6-separate-estimator}
%\end{equation}
%where the correction factors $\frac{1}{m}$ and $\frac{1}{1-m}$ ensures he unbiasedness:
%\begin{equation}
%\begin{split}
%&\mathop{\mathbb{E}}_{\mathbf{u}_i\sim\mathbf{p}^c_{\mathbf{u}_i},\omegai\sim\mathbf{p}^c_{\omegai},\omegai'\sim p_{\mathbf{f}}}[
%m \times w_{path}(\omegai')\frac{1}{m} \frac{\mathbf{f}V}{p_{\mathbf{f}}}\capl_e(\xxo',-\omegai')\\
%&+(1-m) \times w_{path}(\omegai)\frac{1}{1-m} \frac{\capf_{\bm{\theta}}}{\mathbf{p}^c_{\mathbf{u}_i}\mathbf{p}^c_{\omegai}}\capl_e(\xxi',-\omegai)
%]\\
%&= 
%\mathop{\mathbb{E}}_{\omegai'\sim p_{\mathbf{f}}}[
%w_{path}(\omegai')\frac{\mathbf{f}V}{p_\mathbf{f}}\capl_e(\xxo',-\omegai')
%]\\
%&+\mathop{\mathbb{E}}_{\mathbf{u}_i\sim\mathbf{p}^c_{\mathbf{u}_i},\omegai\sim\mathbf{p}^c_{\omegai}}[w_{path}(\omegai)\frac{\capf_{\bm{\theta}}}{\mathbf{p}^c_{\mathbf{u}_i}\mathbf{p}^c_{\omegai}}\capl_e(\xxi',-\omegai)
%]\\
%&=\capl_{path} + \capl_{path}'.
%\end{split}
%\end{equation}
%\paragraph{Choice of selection probability $m$.}
We want to avoid choosing direct-path samples that are occluded and thus contribute nothing to $\capl$.
Therefore, $m$ is defined to be proportional to $V$ by $m\!=\!\tfrac{V\mathbf{f}^c/p_\mathbf{f}}{V\mathbf{f}^c/p_\mathbf{f}+\capf_{\bm{\theta}}^c/\mathbf{p}_{\omegai}^c}$,
so that samples with zero self-visibility are never selected.

\section{Implementation Details}
\begin{algorithm}[tbp]
%\SetAlgoNoLine
\KwIn{
random color channel $c$,
uniform sphere samples $\xxo$,
cosine-hemisphere samples $\omegao$
}
\KwOut{$\xxo,\omegao,\mathbf{u}_i,\omegai,\bm{\beta}_i$, valid mask $m$}
\Comment{sample random outgoing configuration}
$\omegao=\FuncCall{TangentFrame}{$\xxo$}\omegao;\;\omegai=-\omegao$\\
$\xxo, \var{intersect}=\FuncCall{SurfaceIntersect}{$\xxo$,$\omegai$}$\\
\If{not intersect}{
$m=\var{false}$\\
\Return\\
}
$\bm{\beta}_i=1;\;\var{bounce}=1;\;\bx=\xxo;\;m=\var{true}$\\
\Comment{sample incident configuration}
\While{intersect}{
\If{\FuncCall{OnSurface}{$\bx$}}{
\Comment{sample surface scattering}
$\omegai,\bm{\beta}_i=\FuncCall{BSDFSampling}{$\bx$,$-\omegai$,$\bm{\beta}_i$,$c$}$\\
$\xxi=\bx;\;\var{bounce}=\var{bounce}+1$\\
}
\eIf{\FuncCall{InVolume}{$\bx$,$\omegai$}}{
\Comment{sample volume scattering}
$t,\bm{\beta}_i=\FuncCall{MediumSampling}{$\bx$,$\omegai$,$\bm{\beta}_i$,$c$}$\\
$\var{scatter},\bm{\beta}_i=\FuncCall{NullScattering}{$\bx,\omegai,\bm{\beta}_i,c$}$\\
$\bx = \bx + \bo$\\
\If{\var{scatter}}{
$\omegai,\bm{\beta}_i=\FuncCall{PhaseSampling}{$\bx,\omegai,\bm{\beta}_i,c$}$\\
$\xxi=\bx;\;\var{bounce}=\var{bounce}+1$
}
}{
$\bx,\var{intersect}=\FuncCall{SurfaceIntersect}{$\bx,\omegai$}$\\
}
\Comment{throughput-based Russian Roulette}
$\var{terminated},\bm{\beta}_i=\FuncCall{RussianRoulette}{$\bm{\beta}_i,\var{bounce}$}$\\
\If{\var{terminated}}{
$\bm{\beta}_i=0;\;\var{intersect}=\var{false}$\\
}
}
\Comment{reparameterize $\xxi$ to bounding geometry proxy}
$\mathbf{u}_i=\FuncCall{BoundingGeometryIntersect}{$\xxi,\omegai$}$\\
\If{apply direct separation and $\var{bounce}==1$}{
$\bm{\beta}_i=0$\Comment{separate direct scattering}
}
%\Return
\caption{Training data generation}
\label{alg:6-sampling-data}
\end{algorithm}
\begin{algorithm}[tbp]
%\SetAlgoNoLine
$\xxo,\omegao,\mathbf{u}_i,\omegai,\bm{\beta}_i,m\sim$ \var{buffer}\\
$\xxo=\xxo[m];\;\omegao=\omegao[m];\;\mathbf{u}_i=\mathbf{u}_i[m];\;\omegai=\omegai[m]$\\
$\bm{\beta}_i=\bm{\beta}_i[m]$\\
\Comment{scattering distribution loss}
$\var{loss}=\FuncCall{Mean}{$-\bm{\beta}_i\log\mathbf{p}_{\bm{\theta}}(\mathbf{u}_i,\omegai|\xxo,\omegao)$}$\\
\Comment{albedo loss}
$\var{loss}=\var{loss}+\FuncCall{Mean}{$(\bm{\alpha}_{\bm{\theta}}(\xxo,\omegao)-\bm{\beta}_i)^2$}$\\
Take gradient descent step on $\nabla_{\bm{\theta}}\var{loss}$\\
\caption{Training step}
\label{alg:6-training}
\end{algorithm}
\begin{algorithm}[tbp]
%\SetAlgoNoLine
\KwIn{
camera ray ($\xxi$, $\omegai$),
random color channel $c$
}
\KwOut{pixel radiance $\capl$}
$\capl=0;\;\bm{\beta}=0;\;p_{path}=0;\;\var{intersect}=\var{true}$\\
\While{\var{intersect}}{
$\xxo,\var{intersect}=\FuncCall{SurfaceIntersect}{$\xxi,\omegai$};\;\omegao=-\omegai$\\
\Comment{contribution of previous path sampling}
$\capl_e=\FuncCall{Emission}{$\xxo,\omegao$}$\\
$w_{path}=\FuncCall{PowerHeuristic}{$p_{path},\FuncCall{EmitterPdf}{$\xxi$,$\omegai$}$}$\\
$\capl=\capl+w_{path}\bm{\beta}\capl_e$\\
\Comment{emitter sampling (direct scattering)}
$\omegai,p_{emit},\capl_e=\FuncCall{EmitterSampling}{$\xxo$}$\\
$w_{emit}=\FuncCall{PowerHeuristic}{$p_{emit},\FuncCall{BSDFPdf}{$\xxo,\omegai$}$}$\\
$\capl=\capl+w_{emit}\bm{\beta}\frac{\text{bsdf}(\xxo,\omegao,\omegai)}{p_{emit}}\text{visibility}(\xxo,\omegai)\capl_e$\\
\If{\FuncCall{IsNeuralAsset}{$\xxo$}}{
\Comment{emitter sampling (neural asset)}
$\mathbf{u}_i,p_\mathbf{u}\sim \mathbf{p}_{\bm{\theta}}^c(\mathbf{u}_i|\xxo,\omegao)$\\
$\omegai,p_{emit},\capl_e=\FuncCall{EmitterSampling}{$\mathbf{u}_i$}$\\
$\xxi,\_=\FuncCall{SurfaceIntersect}{$\mathbf{u}_i,-\omegai$}$\\
$w_{emit}=\FuncCall{PowerHeuristic}{$p_{emit},\mathbf{p}^c_{\bm{\theta}}(\omegai|\mathbf{u}_i,\xxo,\omegao)$}$\\
$\capl=\capl+w_{emit}\bm{\beta}\frac{\capf_{\bm{\theta}}(\xxo,\omegao,\mathbf{u}_i,\omegai)}{p_\mathbf{u} p_{emit}}\text{visibility}(\xxi,\omegai)\capl_e$\\
}
\Comment{path sampling (direct scattering)}
$\omegai,p_{path}=\FuncCall{BSDFSampling}{$\xxo,\omegao$};\;\mathbf{b}=\frac{\text{bsdf}(\xxo,\omegao,\omegai)}{p_{path}}$\\
$\xxi=\xxo$\\
\If{\FuncCall{IsNeuralAsset}{$\xxo$}}{
\Comment{path sampling (neural asset)}
$\bo'_i,p'_{path}\sim\mathbf{p}^c_{\bm{\theta}}(\bo'_i|\mathbf{u}_i,\xxo,\omegao);\;\mathbf{b}'=\frac{\capf_{\bm{\theta}}(\xxo,\omegao,\mathbf{u}_i,\bo'_i)}{p_\mathbf{u}p_{path}'}$\\
$\xxi',\_=\FuncCall{SurfaceIntersect}{$\mathbf{u}_i,-\omegai'$}$\\
$\mathbf{b}=\mathbf{b}\times \text{self\_visibility}(\xxo,\omegai);\;m=\frac{\mathbf{b}^c}{\mathbf{b}^c+\mathbf{b'}^c}$\\
\Comment{select direct/indirect lobe}
\eIf{$u\sim\text{uniform}(0,1)>m$}{
$p_{path}=p'_{path};\;\mathbf{b}=\mathbf{b}'/(1-m);\;\xxi=\xxi';\;\omegai=\omegai'$\\
}{
$\mathbf{b}=\mathbf{b}/ m$\\
}
}
$\bm{\beta}=\bm{\beta}\times \mathbf{b}$ and apply Russian Roulette\\
}
\caption{Path tracing with our model}
\label{alg:6-inference}
\end{algorithm}
\paragraph{Pseudocode.}
\cref{alg:6-sampling-data} and \ref{alg:6-training} show the pseudocode for generating ground truth path samples and each optimization step during our training.
Note that there is no need to compute the projection factors $\left\vert\frac{
\mathbf{n}_{\xxi}\cdot\omegai
}{
\mathbf{n}_{\mathbf{u}_i}\cdot\omegai
}\right\vert$ and $\frac{\vert\mathbf{n}_{\mathbf{u}_i}\cdot \mathbf{u}_i\vert}
{(\sqrt{\mathbf{u}_i\cdot\mathbf{u}_i})^3}$ in Eqs.~(9) an (10) of the paper during training as 
their log-gradients with respect to training parameters $\bm{\theta}$ are zero.
%and during inference the Jacobian determinants are canceled out as discussed in %\cref{subsec:6-pathtracing}.
\cref{alg:6-inference} shows the pseudocode of our path tracing described in \cref{subsec:6-pathtracing}.
%each path-tracing iteration starts by evaluating the path-sampling contribution of previous iteration followed by the emitter sampling of direct (BSDF) and indirect (neural asset) light transport;
%direct and indirect path samples then are drawn and sotchastically selected as the next-bounce ray.

\paragraph{Baseline rendering.}
The far-field model uses the standard emitter-path sampling at location $\xxo$.
We apply the same lobe selection strategy as above,
and if the indirect scattering is selected, we ignore the intersection tests in next bounce until the ray leaves the asset.

\paragraph{Credit of scenes.}
\begin{itemize}
    \item \textit{Candle}, \textit{Seal}, \textit{Teaset}: Sketchfab.
    \item \textit{Cat}, table textures in  \textit{Cat}, \textit{Seal}, \textit{Candle}: Poly Haven.
    \item \textit{CurlHair}: \citet{resources16}.
    \item \textit{Hair}: Blender Demo.
    \item \textit{Cloud}: \citet{kallweit2017deep}.
    \item \textit{Flowers}: \citet{mullia2024rna}.
    \item \textit{Lego}, environment map, and albedo volume of \textit{Cat}: Mitsuba 3 gallery.
\end{itemize}

\section{Additional Results}
\begin{figure*}[t]
    \centering
    \includegraphics[width=0.99\linewidth]{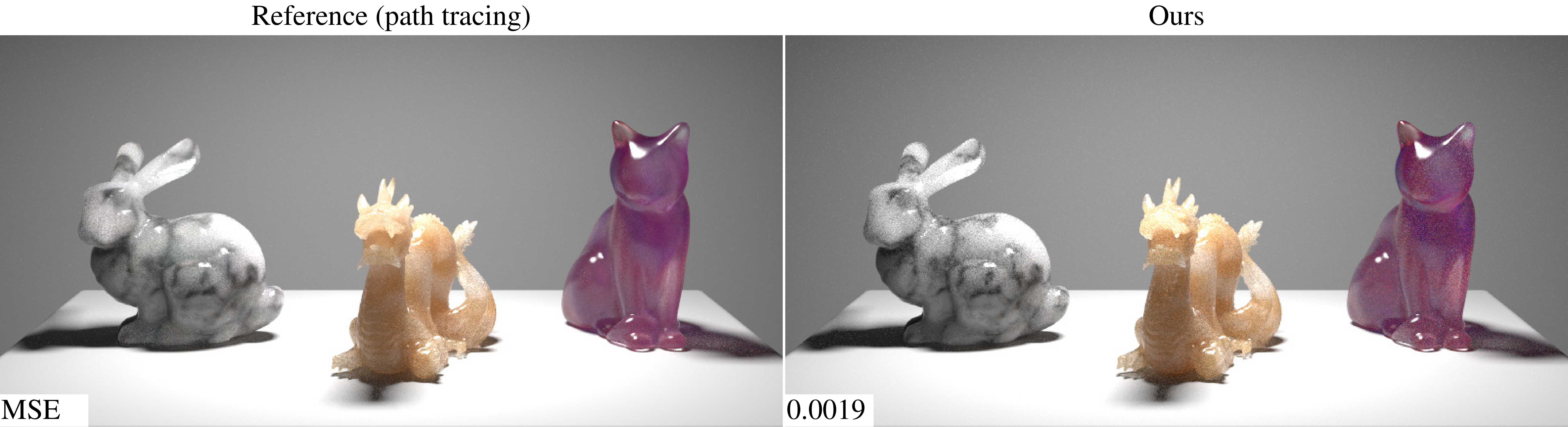}
    \caption{
    \textbf{Scene with composition of neural assets.}
    \textit{Bunny}, \textit{Dragon}, and \textit{Cat} are rendered as three separate neural assets.
    %Our learned neural assets (\textit{Bunny}, \textit{Dragon}, and \textit{Cat}) can be naturally combined into a larger scene, producing rendering consistent with the path-traced reference.
    }
    \label{fig:6-combined}
\end{figure*}
\paragraph{Combination of neural assets.}
As shown in \cref{fig:6-combined}, our learned neural assets can be naturally combined into a larger scene, producing rendering consistent with the path-traced reference.

\begin{figure}
    \centering
    \includegraphics[width=0.99\linewidth]{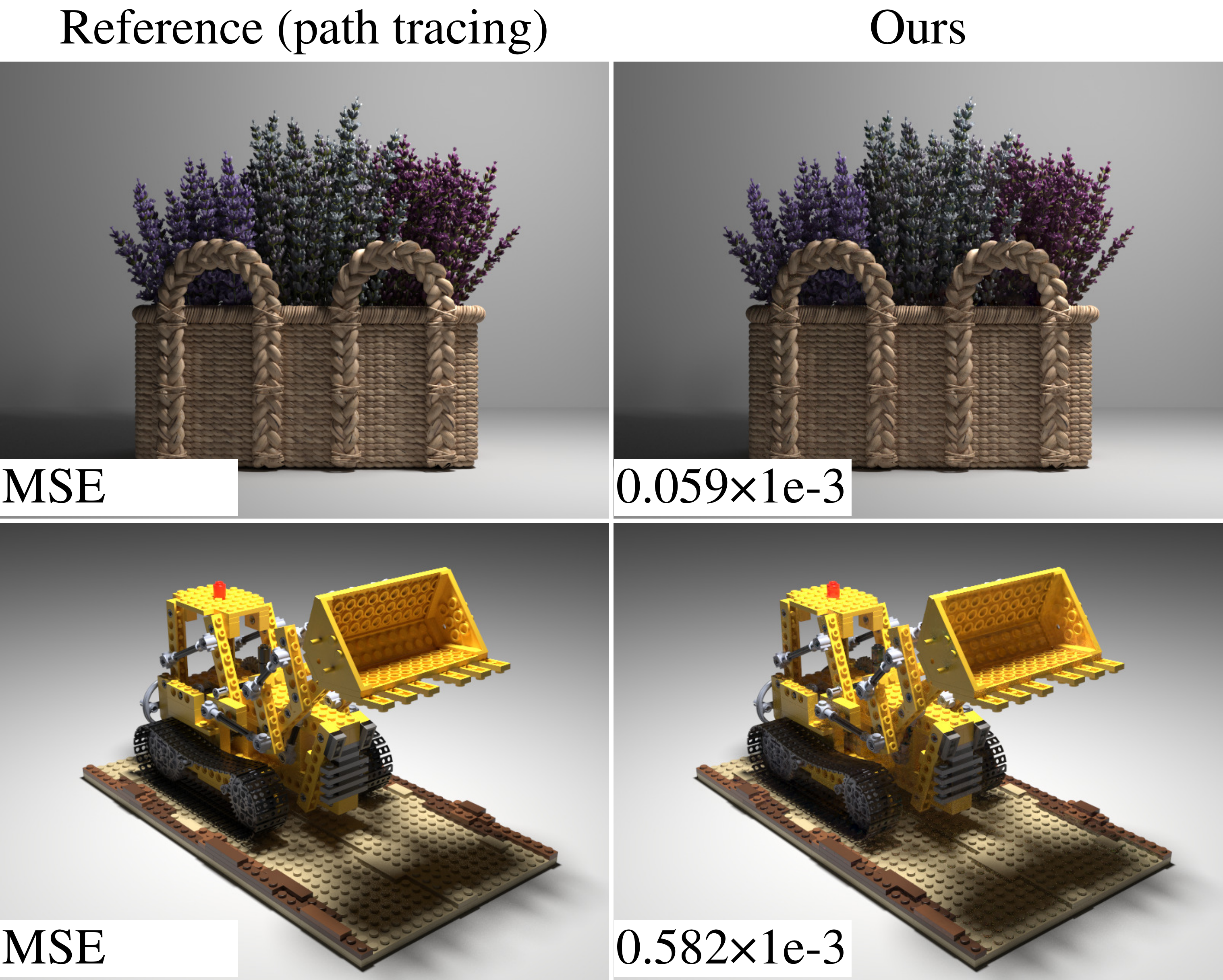}
    \caption{
    \textbf{Assets with complex geometry.}
    1st row: \textit{Flowers}; 2nd row: \textit{Lego}.
    %Our method is able to reconstruct assets with complex geometries such as \textit{Flowers} (first row) and \textit{Lego} (second row).
    }
    \label{fig:6-complex-geometry}
\end{figure}
\paragraph{Complex geometry.}
Figure.~\ref{fig:6-complex-geometry} demonstrates that our method can reconstruct assets with more complex geometry.
To accommodate the increased spatial complexity, we use a $256\times256$ triplane to encode $\xxo$.

\begin{figure*}[t]
    \centering
    \includegraphics[width=0.97\linewidth]{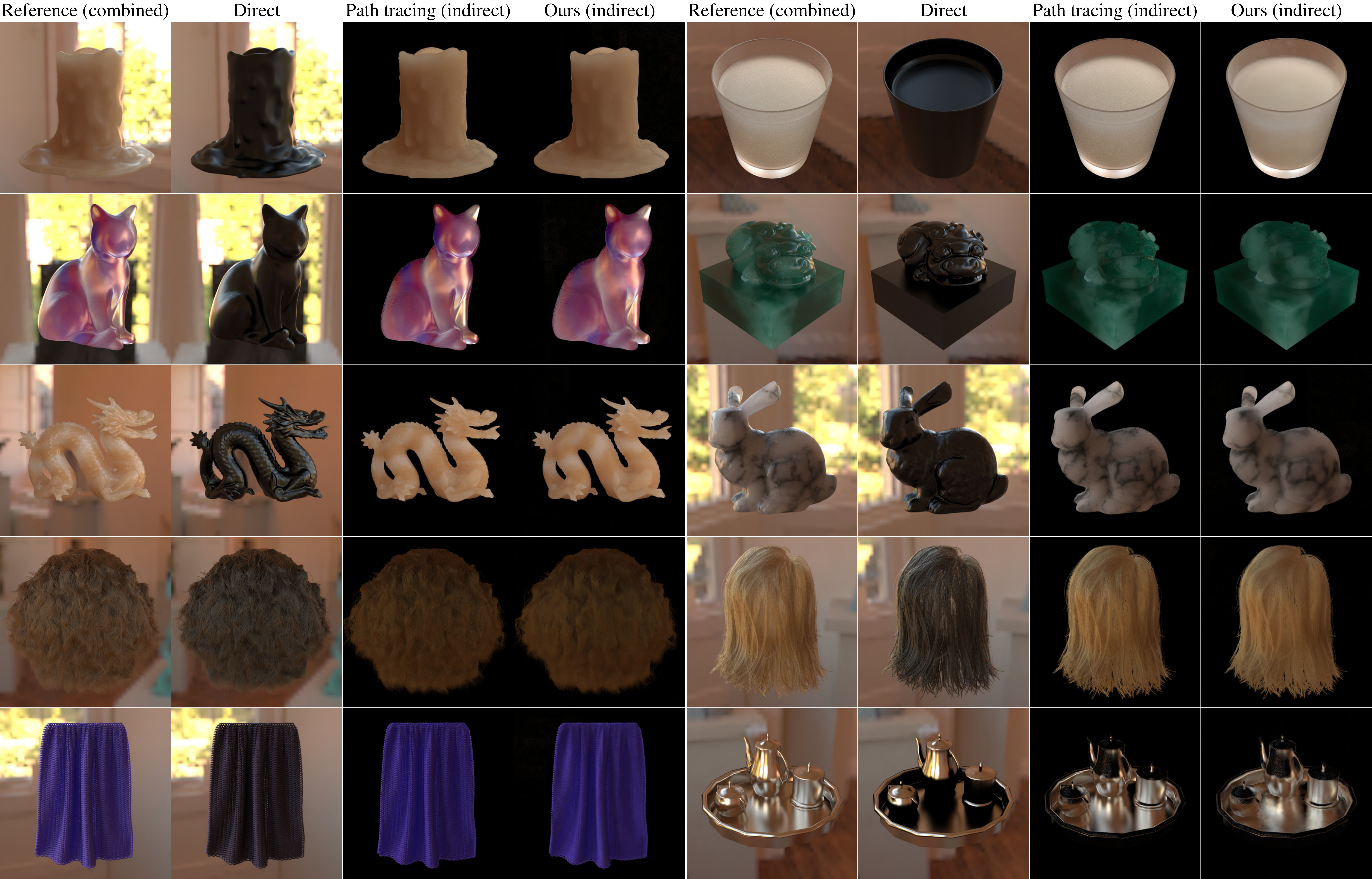}
    \caption{
    \textbf{Visualization of indirect light transport.}
    All assets are rendered under an environment map.
    Our indirect light transport accounts for most of the visible illumination for volumetric assets, and the contribution of indirect illumination remains noticeable for fiber- and surface-based assets.
    }
    \label{fig:6-indirect}
\end{figure*}
\paragraph{Visualization of direct and indirect light transport.}
We further examine the contribution of the learned indirect light transport to the final renderings in \cref{fig:6-indirect}.
As discussed in the main paper, 
the direct illumination for volumetric assets is mainly the specular highlight,
while our indirect model accounts for most of the visible scattering effects.
The fiber rendering exhibits a stronger direct component but looks gray without the indirect scattering.
For surface-based assets, the direct term contributes more to the final appearance,
but indirect illumination remains essential for capturing effects such as interreflection in the \textit{Teaset}.
Across all the assets, our indirect scattering contributes a substantial portion of the total light transport.

\begin{figure*}[t]
    \centering
    \includegraphics[width=0.99\linewidth]{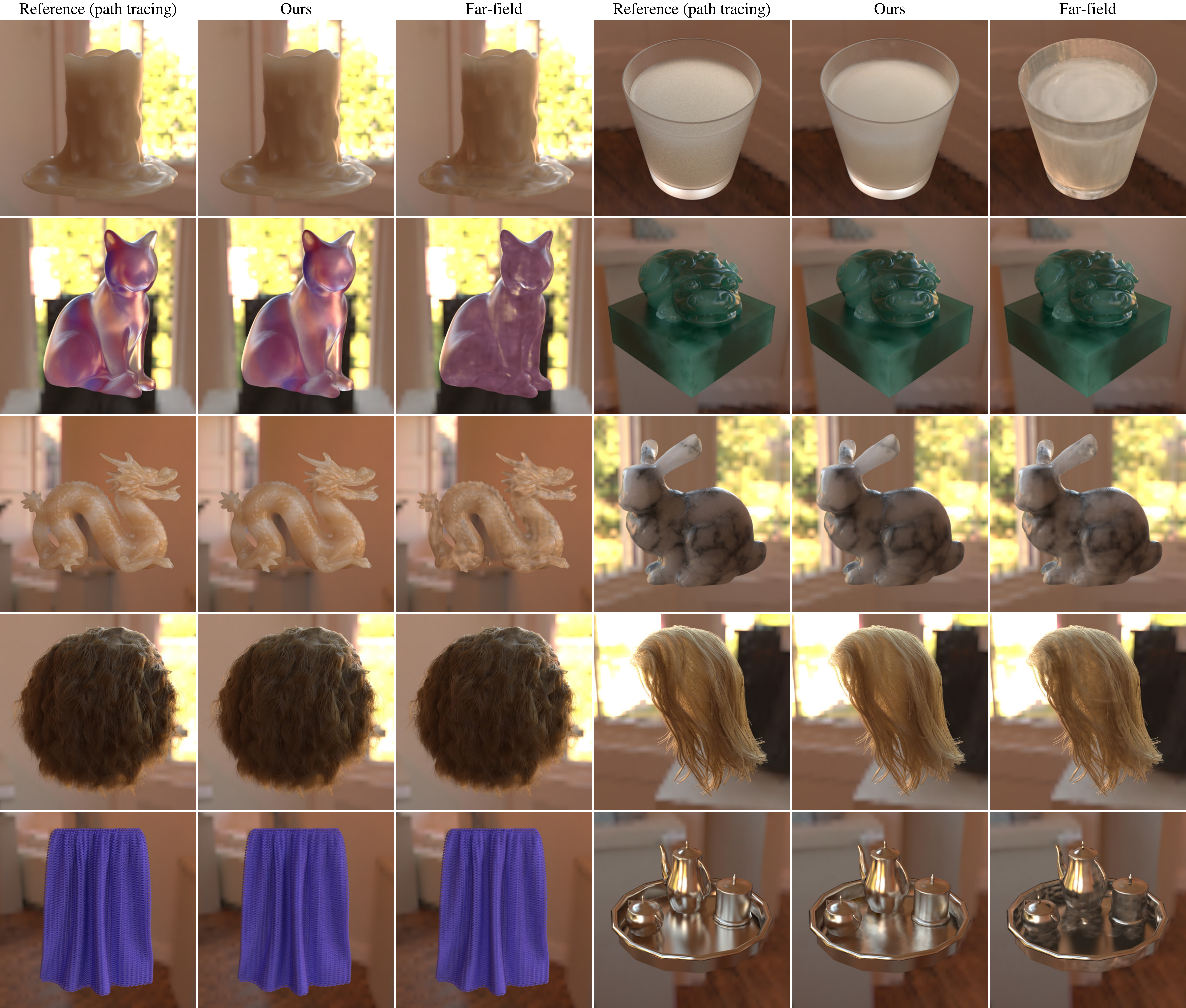}
    \caption{\textbf{Qualitative rendering comparison under far-field illumination.}
    We use an environment light without any background geometries for each scene.
    All images are rendered using 2048 spp.
    Note \methodA{} fails to converge on \textit{Milk}, \textit{Cat}, and \textit{Teaset} dues to its high optimization variance.
    }
    \label{fig:6-farfield}
\end{figure*}
\paragraph{Comparison under far-field lighting.}
We also evaluate each asset under a pure environment light (\cref{fig:6-farfield}), 
a configuration that matches the far-field assumption of the baseline.
It can be seen both the far-field and our model closely reproduce the path-traced light transport, except for \textit{Milk}, \text{Cat}, and \textit{Teaset}, where far-field training fails to converge due to the high variance of the regression loss.
The corresponding quantitative results are provided in \cref{tab:6-farfield}.

\begin{figure*}[t]
    \centering
    \includegraphics[width=0.99\linewidth]{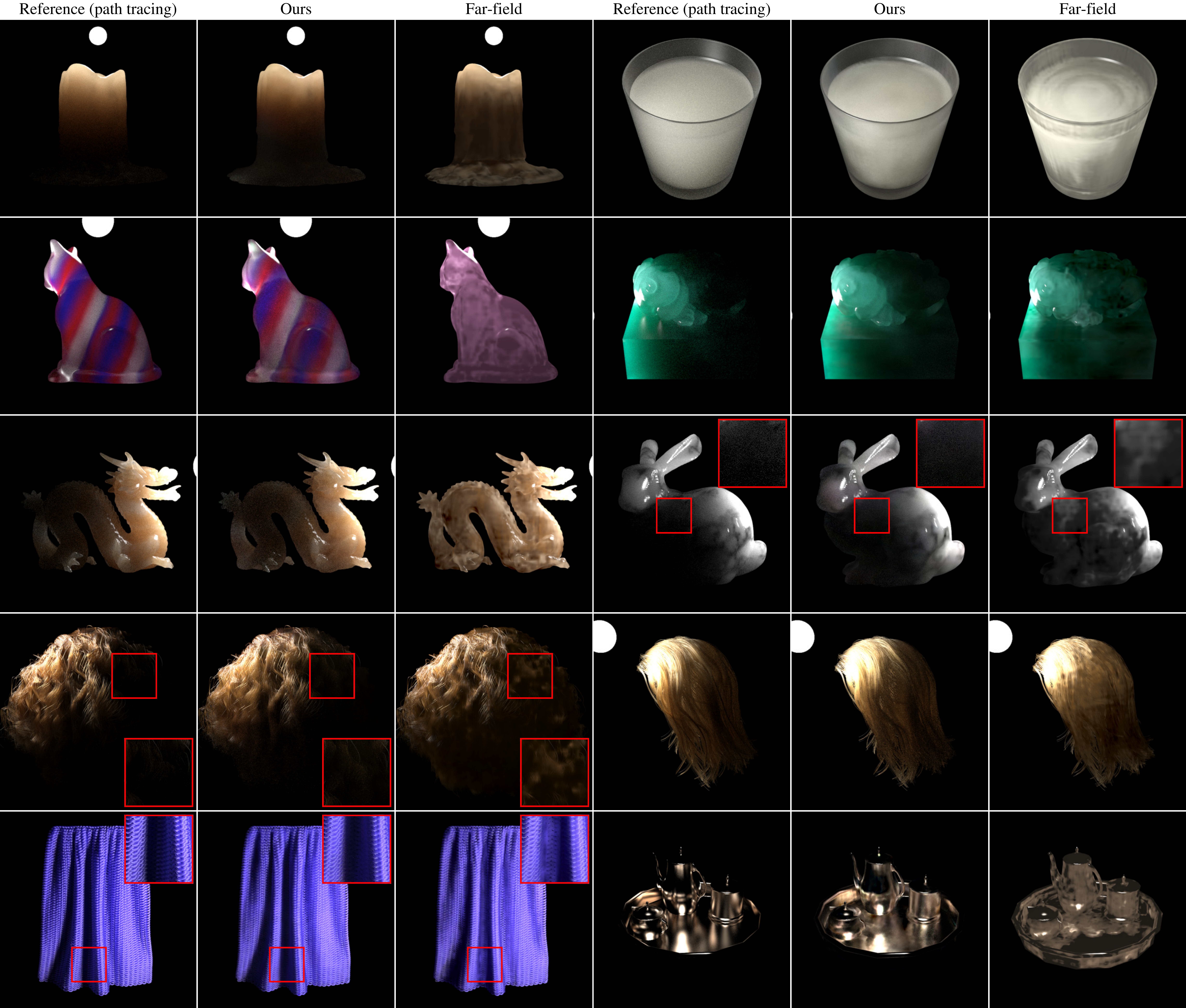}
    \caption{\textbf{Qualitative rendering comparison under a near-field area light.}
    We use a sphere area light placed near the asset without any background geometries. All images are rendered using 8192 spp except for \textit{Milk}, \textit{Seal}, and \textit{Dragon} where the path tracing uses 16384 spp.
    It can be seen that \methodA{} overestimates incoming radiance in regions occluded from the emitter, producing overly bright appearances.
    Normalizing flows are inefficient at reconstructing high-frequency signals, so our interreflection on \textit{Teaset} appears oversmoothed under a sharp near-field light.
    }
    \label{fig:6-nearfield}
\end{figure*}
\paragraph{Comparison under a single area light.}
Figure~\ref{fig:6-nearfield} and \cref{tab:6-nearfield} compare the methods under a single sphere area light.
Without background geometry to bounce light, the incident illumination varies more sharply in space, which amplifies the error of the far-field model.
Our method reconstructs the appearance well overall but oversmooths the high-frequency details on \textit{Teaset}.
This is caused by the limited expressivity of normalizing flow discussed in Sec.~4.4 of the paper.
\begin{table}[t]
    \centering
    \setlength\tabcolsep{2 pt}
    \caption{\textbf{Rendering error under an environment light.}
    Both our model and far-field baseline correctly captures the light transport in most of the assets. Farfield fails to reconstruct \textit{Milk}, \textit{Cat}, and \textit{Teaset} due to the high optimization variance.
    }
    \resizebox{0.99\linewidth}{!}{
    \begin{tabular}{l c c c c c c c c c c}
        \toprule
        \multirow{2}{*}{\textbf{Method}} 
        &
        Candle & Milk & Cat & Seal & Dragon & Bunny & CurlHair & Hair & Fabric & Teaset\\
        &\multicolumn{10}{c}{\textbf{100$\times$MSE$\downarrow$}}\\
        \midrule    
\textbf{\methodA{}} & 0.013 & 0.598 & 1.657 & 0.004 & 0.058 & 0.031 & 0.008 & 0.052 & 0.035 & 0.724\\
\textbf{Ours} & 0.010 & 0.079 & 0.153 & 0.003 & 0.018 & 0.011 & 0.006 & 0.033 & 0.026 & 0.092\\
        \bottomrule
    \end{tabular}
    }
    \label{tab:6-farfield}
\end{table}
\begin{table}[t]
    \centering
    \setlength\tabcolsep{2 pt}
    \caption{\textbf{Rendering error under a single area light.}
    Our method exhibits less reconstruction error overall.
    }
    \resizebox{0.99\linewidth}{!}{
    \begin{tabular}{l c c c c c c c c c c}
        \toprule
        \multirow{2}{*}{\textbf{Method}} 
        &
        Candle & Milk & Cat & Seal & Dragon & Bunny & CurlHair & Hair & Fabric & Teaset\\
        &\multicolumn{10}{c}{\textbf{100$\times$MSE$\downarrow$}}\\
        \midrule    
\textbf{\methodA{}} & 0.092 & 1.178 & 1.013 & 0.155 & 1.180 & 0.216 & 0.201 & 0.767 & 1.080 & 14.240\\
\textbf{Ours} & 0.052 & 0.152 & 0.551 & 0.120 & 0.492 & 0.100 & 0.160 & 0.566 & 0.871 & 2.389\\
        \bottomrule
    \end{tabular}
    }
    \label{tab:6-nearfield}
\end{table}

\begin{figure*}[t]
    \centering
    \includegraphics[width=0.99\linewidth]{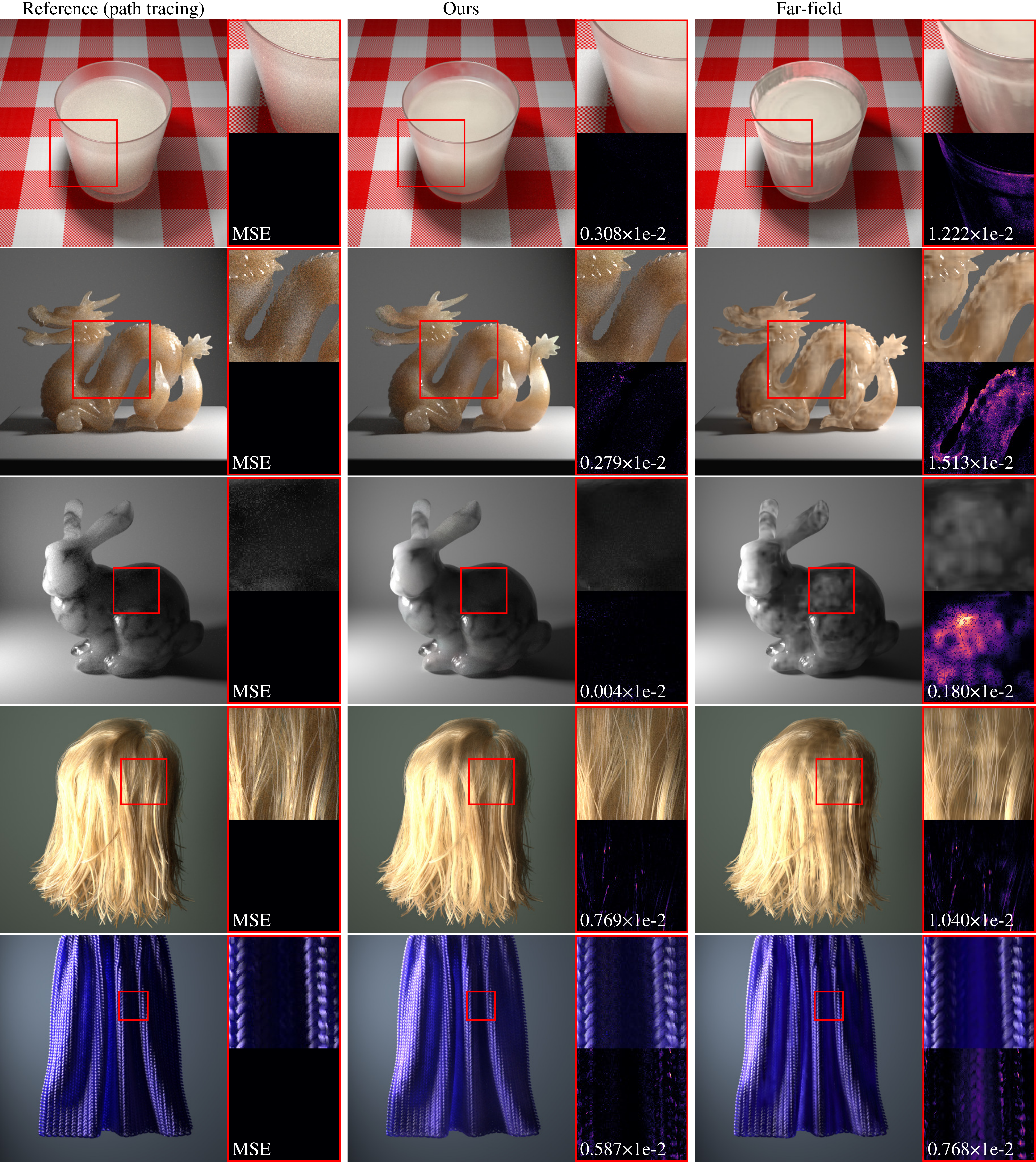}
    \caption{\textbf{Additional rendering comparison.}}
    \label{fig:6-acc}
\end{figure*}

\begin{figure*}[t]
    \includegraphics[width=0.97\linewidth]{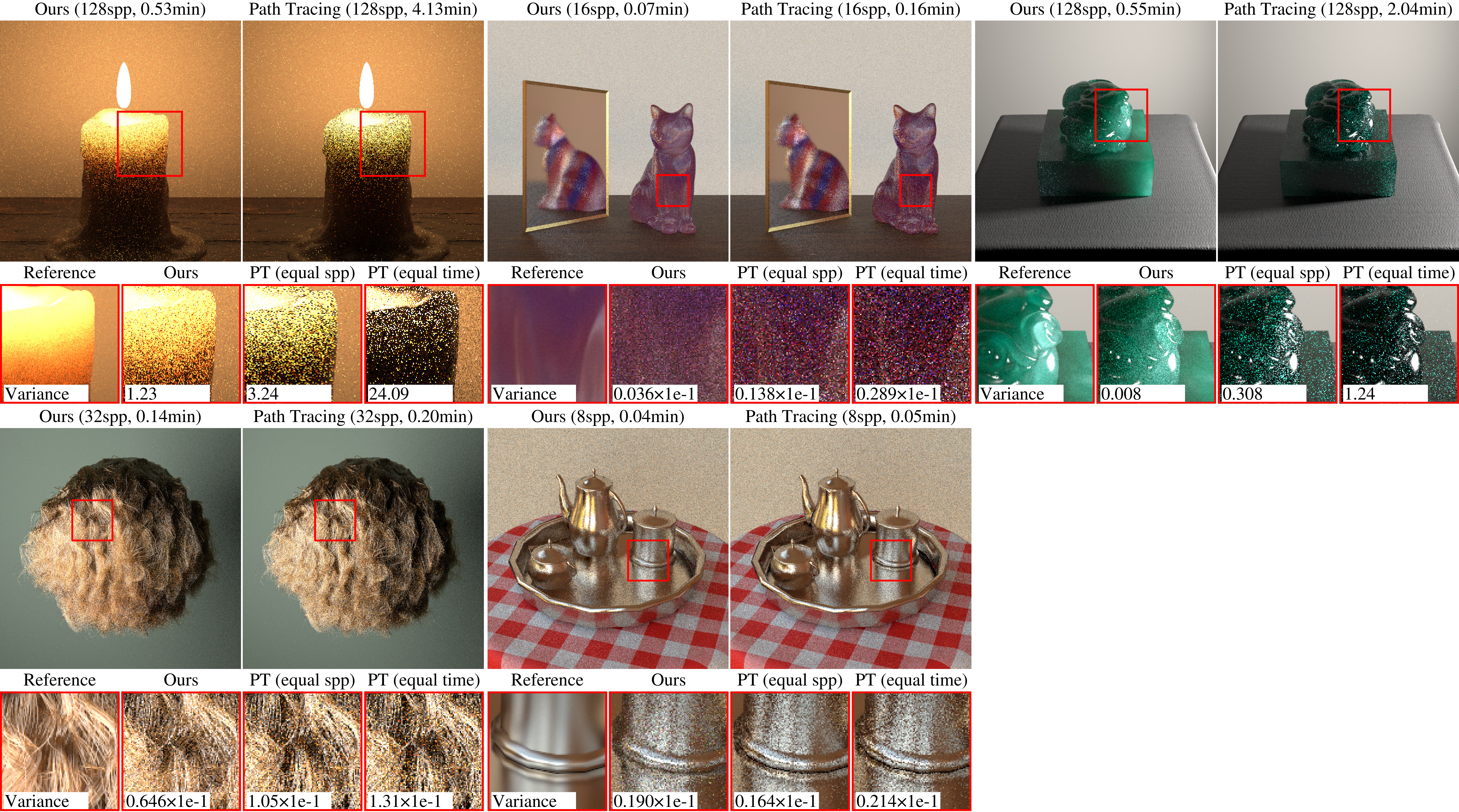}
    \caption{\textbf{Additional rendering variance comparison.}
    }
    \label{fig:6-efficiency}
\end{figure*}

\paragraph{Additional rendering and variance comparison.}
Figures~\ref{fig:6-acc} and \ref{fig:6-efficiency} show the rendering and variance comparison for the rest of the scenes in the experiments of the paper.

\paragraph{GPU memory usage.}
Since we generate training data online,
the GPU memory required during training depends on asset complexity,
which is around 6.8 GB for the volumetric assets, 5.3 GB for \textit{Teaset}, and 7.2-11.4 GB for fiber-based assets (\textit{CulrHair}, \textit{Hair}, \textit{Fabric}).
The network optimization uses around 4.5 GB GPU memory excluding the dataset generation.
Table~\ref{tab:6-memory} reports the GPU memory usage during inference.
Overall, all methods use similar GPU memory, but our method consumes slightly more memory due to the usage of more neural networks.
\begin{table}[t]
    \centering
    \setlength\tabcolsep{2 pt}
    \caption{\textbf{GPU memory usage for rendering \cref{fig:6-efficiency} and Fig.~14 of the paper.}
    Compared to standard path tracing,
    \methodA{} and our model requires additional memory for neural networks,
    but this overhead is small.
    Our model uses more memory due to more neural network evaluations.
    }
    \resizebox{0.99\linewidth}{!}{
    \begin{tabular}{l c c c c c c c c c c}
        \toprule
        \multirow{2}{*}{\textbf{Method}} 
        &
        Candle & Milk & Cat & Seal & Dragon & Bunny & CurlHair & Hair & Fabric & Teaset\\
        &\multicolumn{10}{c}{\textbf{GB$\downarrow$}}\\
        \midrule    
        \textbf{\methodA{}} & 3.6 & 4.3 & 3.7 & 3.8 & 3.8 & 3.7 & 8.2 & 3.7 & 5.8 & 4.3\\
\textbf{PT} & 3.3 & 4.2 & 3.4 & 3.7 & 3.6 & 3.5 & 8.0 & 3.5 & 5.0 & 4.0\\
\textbf{Ours} & 3.7 & 4.6 & 3.8 & 4.2 & 3.7 & 3.8 & 8.8 & 4.2 & 6.6 & 4.6\\
        \bottomrule
    \end{tabular}
    }
    \label{tab:6-memory}
\end{table}

\clearpage
\newpage
\end{appendix}

\end{document}